\documentclass[a4paper,fleqn,usenatbib]{mnras}

\usepackage{newtxtext,newtxmath}

\usepackage[T1]{fontenc}
\usepackage{ae,aecompl}

\usepackage{graphicx}	

\usepackage{amsmath}	
\usepackage{amssymb}	
\usepackage{epstopdf}	

\usepackage{soul}
\usepackage{pifont}

\usepackage{comment}

\usepackage{scalerel,tikz}
\usetikzlibrary{svg.path}
\definecolor{orcidlogocol}{HTML}{A6CE39}
\tikzset{orcidlogo/.pic={
 \fill[orcidlogocol] svg{M256,128c0,70.7-57.3,128-128,128C57.3,256,0,198.7,0,128C0,57.3,57.3,0,128,0C198.7,0,256,57.3,256,128z};
 \fill[white] svg{M86.3,186.2H70.9V79.1h15.4v48.4V186.2z}
 svg{M108.9,79.1h41.6c39.6,0,57,28.3,57,53.6c0,27.5-21.5,53.6-56.8,53.6h-41.8V79.1z M124.3,172.4h24.5c34.9,0,42.9-26.5,42.9-39.7c0-21.5-13.7-39.7-43.7-39.7h-23.7V172.4z}
 svg{M88.7,56.8c0,5.5-4.5,10.1-10.1,10.1c-5.6,0-10.1-4.6-10.1-10.1c0-5.6,4.5-10.1,10.1-10.1C84.2,46.7,88.7,51.3,88.7,56.8z};
}}
\newcommand\orcid[1]{\href{https://orcid.org/#1}{\mbox{\scalerel*{
\begin{tikzpicture}[yscale=-1,transform shape]
\pic{orcidlogo};
\end{tikzpicture}
}{|}}}}

\title[AGN Feedback in high-z galaxies]
{On the rapid growth of SMBHs in high-z galaxies: the aftermath of Population III.1 stars}

\author[Sanati et al.]{Mahsa Sanati$^{1,2}$\thanks{E-mail: mahsa.sanati@physics.ox.ac.uk},
Julien Devriendt$^{1}$,
Sergio Martin-Alvarez$^{3}$,
Adrianne Slyz$^{1}$,
Jonathan C. Tan$^{2,4}$
\\
$^{1}$Department of Physics, University of Oxford, Keble Road, Oxford OX1 3RH, UK\\
$^{2}$Department of Space, Earth \& Environment, Chalmers University of Technology, SE-412 96 Gothenburg, Sweden\\
$^{3}$Kavli Institute for Particle Astrophysics \& Cosmology (KIPAC), Stanford University, Stanford, CA 94305, USA\\
$^{4}$Department of Astronomy, University of Virginia, Charlottesville, VA 22904-4235, USA\\
}

\date{Accepted XXX. Received YYY; in original form ZZZ}

\pubyear{2025}

\begin{document}

\newcommand{\papertitle}[2]{\textbf{#1 \citep{#2}}.\newline}

\label{firstpage}
\pagerange{\pageref{firstpage}--\pageref{lastpage}}
\maketitle

\begin{abstract}
%
The vast amount of energy released by active galactic nuclei (AGN) is increasingly recognized as a key driver of evolution not only in massive galaxies and clusters, but also in low-mass dwarf galaxies.
Despite this, their role in the early stages of galaxy formation and in self-regulating the rapid growth of the first and abundant supermassive black holes (SMBHs) remains poorly understood. 
Through new high-resolution zoom-in cosmological simulations, we follow the co-evolution of $10^{5}\,M_{\odot}$ black hole seeds with their host galaxy. 
%
%
%
By modelling ionizing feedback from a Pop III.1 progenitor, we capture often-neglected preheating effects, making our results applicable to a wide range of SMBH formation scenarios, including in internally or externally irradiated halos.
%
%
%
%
The simulated suite progressively spans physics ranging from no AGN feedback and Eddington-limited thermal feedback, to more complex setups including non-Eddington-limited thermal, kinetic and radiative feedback {--} explored for both low and enhanced AGN power.
Across all our models, we find that black hole seeds efficiently reach masses of $
\sim10^{7}\,M_{\odot}$ by $z = 8$. Although they exhibit notably different mass growth histories, these latter seem unimpeded by the presence of AGN feedback. 
%
%
The simulation including radiative feedback is the most distinct, with super-Eddington episodes driving fast and mass-loaded gas outflows (exceeding $2500\,\mathrm{km\,s^{-1}}$) up to $\sim50\,\mathrm{kpc}$, along with minor stellar mass suppression in the host galaxy.  
%
Our measurements are in broad agreement with moderate luminosity quasars recently observed by JWST, producing overmassive black holes (SMBH-to-galaxy mass ratios $0.01 - 1$), dynamical masses of $\sim10^{9.5}\,M_{\odot}$, stellar masses of $\sim10^{8.5}\,M_{\odot}$, and high, though short-lived, Eddington fraction accretion rates.  
These results advocate for a scenario where AGN feedback allows for rapid SMBH growth during the reionisation era, while driving winds that extend deep into the intergalactic medium {--} shaping host galaxies as well as more distant surroundings.
\newline

\end{abstract}

\definecolor{brown}{rgb}{0.5, 0.3, 0.0}
\definecolor{orange}{rgb}{0.8, 0.5, 0.0}



\section{Introduction}\label{sec:intro}









The co-evolution of supermassive black holes (SMBHs) and their host galaxies has been extensively studied in the local Universe \citep[e.g.,][]{kormendy_coevolution_2013,2015ApJ...813...82R, 2020ARA&A..58..257G,2021ApJ...921...36B}
, revealing tight correlations between the observed SMBH masses and various host galaxy properties.
%
%
While these relationships are well established at low
redshift, it remains unclear whether they hold at high redshift, and how they evolve with time \citep[e.g.,][]{2023MNRAS.518.2177G}. There are studies that find observational evidence of redshift evolution \citep[e.g.,][]{2010ApJ...708..137M, 2011ApJ...742..107B, 2023ApJ...943..133F}, others that report consistency with local relations \citep[e.g.,][]{2003ApJ...583..124S, 2015ApJ...805...96S, 2020ApJ...889...32S,2023ApJ...954..173L}, and potential evidence for only a subset of relations to evolve \citep[e.g.,][]{2009ApJ...706L.215J,2020ApJ...888...37D}.
One of the strongest local correlations, that between SMBH mass and the velocity dispersion of the galaxy bulge, is not accessible at high redshift, as isolating the bulge component is challenging due to limited spatial resolution and the irregular morphologies of early galaxies.  
Thus many observational studies focused on the relation between SMBH mass and total stellar mass instead.
These studies, examining smaller and more
distant galaxies, frequently find that black holes in early galaxies appear systematically over-massive relative to their inferred stellar masses \citep{2024A&A...691A.145M, andika_tracing_2024, 2024ApJ...964...90S}. These deviations typically lie $1-2\,\rm{dex}$  above the local scaling relations 
\citep{2024MNRAS.531..355U, harikane_jwstnirspec_2023, Kokorev:2023,  2024Natur.627...59M, 2024Natur.628...57F, 2024Natur.636..594J, 2024ApJ...960L...1N}
%
%
While this offset is subject to selection biases, such as detection preference for massive black holes \citep[e.g.,][]{2024ApJ...969L..18A, 2024A&A...686A.256L}, it suggests that SMBHs may experience rapid growth during the early stages of galaxy formation \citep[e.g.][]{2023MNRAS.526.3250S, pacucci_jwst_2023, 2024arXiv240906772T, 2024arXiv241204983T, 2025arXiv250403551J, 2025arXiv250603354B}.

Observations of quasars hosting massive BHs already in place at redshifts $z\sim4-14$ \citep{2023ApJ...954L...4K, Larson:2023, Kokorev:2023, maiolino_jades_2023, 2024A&A...691A.145M, 2024ApJ...965L..21K,2024NatAs...8..126B,2024ApJ...964...39G} further support this picture, motivating models of SMBH evolution that rely on massive initial seeds \citep{2014MNRAS.443.2410F, banik_formation_2019, 2021JCAP...07..039X, 2023Natur.615..597N, 2025arXiv250400075S} as well as periods of rapid, potentially super-Eddington growth \citep{Lupi:2016, inayoshi_assembly_2020, 2023MNRAS.519.1837S,   2025MNRAS.537.2559H, 2023A&A...670A.180M, 2024MNRAS.527.1033B, 2024ApJ...976...96P}.
%
%
These moderate-luminosity quasars reside within faint, low-mass dwarf-size galaxies, showing significant activity from the AGN. 
%
%
Yet, the impact of these early AGN on both black hole growth and the evolution of their host galaxies remains poorly understood, highlighting the pressing necessity to map out AGN feedback and SMBH-galaxy co-evolution at early cosmic times.
%
%



One of the primary effects associated with AGN feedback on galaxy evolution is the regulation of star formation \citep[e.g.,][]{2023A&A...679A.151M, 2024MNRAS.528.4891G, 2024ApJ...961..163B}. 
This regulatory role has been proposed in high-redshift galaxies as a potential driver of bursty star formation histories and episodes of early quenching or "mini-quenching" \citep{Belli2024, D'Eugenio2024, 2025A&A...697A..88L}, though mainly in galaxies with masses above the dwarf regime.
As observational evidence grows \citep{2017ApJ...846...32D, 2018MNRAS.473.3717F, 2023MNRAS.524.2312E,2024MNRAS.533.1111E, 2020MNRAS.497..698T}, numerous simulations have explored the mechanisms behind this burstiness in the early Universe \citep[e.g.,][]{2022MNRAS.515.2386R, Sun2023, Dome2025, Martin-Alvarez2025}.
%
%

The role of AGN feedback in shaping the low-mass end of the galaxy population is also gaining increasing attention, as ever more compelling observations of local dwarf galaxies exhibiting energetic, galaxy-wide outflows emerge \citep[e.g.,][]{2018MNRAS.476..979P, 2019ApJ...884...54M, 2020ApJ...905..166L, 2022MNRAS.511.4109D, 2023ApJ...950...33A, 2025ApJ...984..155S,2025A&A...697A.235R}.
%
These outflows, often enriched with metals, increasingly point to AGN-driven feedback as a key channel for injecting energy and momentum into the interstellar medium (ISM) and far into the circumgalactic medium (CGM) \citep{2007ApJ...655L..17B, 2015ApJ...811..149C,2015ApJ...809..147H, 2017MNRAS.465.1682H,2024ApJ...960...55Z},  effectively heating and expelling gas from galaxies. 
While supernova feedback {--} particularly when modelled at high resolution \citep[e.g.,][]{2020MNRAS.491.1656A,sanati_how_2023,revaz_compactness_2023,2023ApJ...950..132H,2024MNRAS.528.5412R,2025arXiv250418620L,2025arXiv250418384S,2025ApJ...986..214G,2025arXiv250408035P} and incorporating non-thermal processes such as cosmic rays, radiation, and magnetic fields {--} can generate moderately energetic winds \citep{2007MNRAS.378..385P, 2015MNRAS.451...34R, 2020A&A...638A.123D,2021MNRAS.501.3640H,2021MNRAS.506.2199G, 2022MNRAS.513.5000F,2024MNRAS.530.3617R, martin-alvarez_extragalactic_2023}, it continues to face various shortcomings.  Specific examples include the high-velocity tail of galactic winds \citep{carniani_jades_2024, 2025arXiv250407074S}, as well as the entrainment of cold and molecular gas frequently observed in outflows \citep[e.g.,][]{2020A&ARv..28....2V,2024MNRAS.528.4976D,2025arXiv250601088A,2025arXiv250415357B}. 
As a result, many models have turned to energy input from AGNs to account for such large-scale effects in low mass galaxies \citep[e.g.,][]{2019MNRAS.484.2047K,2025MNRAS.540.1928K,2023ApJ...957...16S,2024A&A...690A.286A}.
%

%
%

One of the central challenges in modeling the AGN-galaxy interplay lies in capturing the vast dynamical range involved, from accretion onto black holes at small scales to large-scale galactic feedback.
Both cosmological and idealized galaxy simulations have been extensively used to investigate the formation and growth of SMBHs.
Large-volume cosmological simulations such as Horizon-AGN \citep{Dubois:2014,2021A&A...653A.154T}, Illustris \citep{Vogelsberger:2013,Genel:2014,Torrey:2014, Sijacki:2015}, EAGLE \citep{Crain:2015, Schaye:2015},  SIMBA \citep{Dave:2019}, and ASTRID \citep{Ni:2022b}
focus on evolving BH populations, in agreement with local observational constraints. 
However, these sophisticated cosmological simulations are limited by numerical resolution and the absence of key physical processes such as radiative transfer and magnetohydrodynamics, which prevent a more realistic treatment of AGN feedback.
Meanwhile, idealized simulations \citep[e.g.,][]{2014ApJ...783...50L,2015MNRAS.454.3445C, beckmann,2023MNRAS.526.3967M} have focused mainly on investigating the small-scale behaviour of BHs. 
%
Zoom-in simulations \citep[e.g.,][]{costa_quenching_2018, 2018MNRAS.473.4003B, 2022MNRAS.513.3768I, 2023MNRAS.520.5394W, 2025arXiv250408041F,2025MNRAS.537.2559H} are used as a complementary technique to overcome the numerical limitations of large-volume simulations. 
They provide the high resolution necessary to form the dark matter (DM) progenitors of SMBHs host halos at high redshifts and allow a detailed tracking of the properties of the gaseous environment where these BHs grow.

Heavy-seed SMBHs are typically modeled in simulations with a mass resolution of $10^4$–$10^5\,M_{\odot}$, using a threshold criterion either based on DM halo mass \citep[e.g.,][]{vogelsberger2014,Schaye:2015} or on gas properties \citep[e.g.,][]{2017MNRAS.470.1121T}. However, the pre-formation thermal and dynamical state of the gas is often overlooked, a simplification that can strongly influence early accretion rates and AGN–galaxy interplay \citep[e.g.,][]{JohnsonBromm:2007, 2023MNRAS.521.2845C}.
In direct collapse black hole (DCBH) scenarios, the host halo is typically pre-heated by an external Lyman–Werner radiation field to suppress molecular cooling and prevent fragmentation, creating the right conditions for massive seed formation \citep[e.g.,][]{begelman_formation_2006}. 
Similarly, in Population III.1 (Pop III.1) star progenitor models 
the massive star ionizes and heats the host halo, modifying its density and temperature before the black hole forms (Sanati et al submitted). 
These effects are accounted for in the present study by self-consistently modelling the formation and the radiative feedback of the Pop III.1 star prior to its collapse into a SMBH.

Pop III.1 stars offer a natural pathway for producing heavy black hole seeds, forming in locally isolated dark matter minihalos \citep[][]{banik_formation_2019,singh_formation_2023,2025MNRAS.536..851C} (see \citet{2024arXiv241201828T} for a review). Protostars in this scenario grow to masses comparable to those of heavy black hole seeds through energy injection from the annihilation of weakly interacting massive particle (WIMP) dark matter \citep{spolyar_dark_2008,2008IAUS..255...24T,2009ApJ...692..574N,2009ApJ...693.1563F,2015ApJ...799..210R}. The injected energy keeps the protostar in an extended, cool state (photospheric temperature $\sim 10^4\,\rm{K}$) and low ionizing feedback, allowing it to accrete a large fraction i.e., up to $\sim 10^5\:M_\odot$, of the baryonic mass contained in the minihalo.
In the cosmological framework of Pop III.1 seeding \citep{banik_formation_2019}, this evolutionary pathway occurs only in isolated, undisturbed minihalos, where slow baryonic contraction allows for adiabatic compression of the dark matter density surrounding the protostar. In contrast, minihalos that remain intact after exposure to ionizing radiation develop elevated free electron fractions, which enhance $\rm H_2$ and HD cooling. This leads to fragmentation into multiple lower-mass stars. Since these so-called Pop III.2 stars are predicted to have masses of only $\sim 10-100\:M_\odot$ \citep[e.g.,][]{2006MNRAS.366..247J}, they are unlikely to be an effective source of SMBHs formation.
%
The consistency of SMBH number densities predicted by the Pop III.1 scenario with observational constraints from both the local and high-z Universe \citep[][Sanati et al submitted, Petkova et al in prep.]{hayes_glimmers_2024, 2025arXiv250117675C} hints at the potentially important role that Pop III.1 sources play in the formation of SMBHs. 



In this manuscript we study the AGN feedback in high-resolution zoom-in cosmological simulations, exploring the co-evolution of the BH with its host galaxy.
Our physically motivated AGN feedback models capture both radio and quasar modes and uniquely include radiative transfer to model AGN radiation effects on the ISM and CGM. 
The simulations achieve parsec-scale resolution in faint, low-mass galaxies at high redshift, allowing for an unprecedentedly detailed examination of 
AGN-driven outflows during uncapped, super-Eddington accretion episodes, and their interaction with the host environment.
We compare our predictions with JWST observations providing new insight into the physical origin of observed outflows, BH–galaxy scaling relations, and the signatures of early AGN feedback.
%
%
This paper is organized as follows. The numerical framework to generate and evolve our simulations is described in detail in \S\ref{sec:meth}. The results are presented in \S\ref{sec:res}. 
Finally, a summary of our main conclusions and a brief discussion is presented in \S\ref{sec:sum}.

\section{Numerical methods and simulations}\label{sec:meth}

We generate all the cosmological 
simulations studied in this work using the \textsc{Ramses} code \citep{Teyssier2002}. In addition to the collisionless dark matter and stellar particles, \textsc{Ramses} employs an adaptive mesh refinement octree grid to solve the evolution of gas. The code
simultaneously and self-consistently 
model radiative transfer \citep{rosdahl_ramses-rt_2013, 2015MNRAS.449.4380R, 2018MNRAS.479..994R, 2022MNRAS.515.2386R} and constrained transport (CT) magneto-hydrodynamics \citep[MHD;][]{2006A&A...457..371F, 2006JCoPh.218...44T},
and black hole formation \citep{Dubois:2010}, 
in addition to treating baryonic physics, such as redshift-evolving and uniform UV heating, gas cooling, star formation, and stellar feedback.
Below we provide a brief summary of essential features in the code.

\begin{table}
\caption{\small
  Parameters varied in the simulation runs. 
  Columns are: 
  1) Model ID; 
  2) AGN feedback components; 
  3) AGN thermal feedback efficiency;
  4) AGN kinetic feedback efficiency;
  5) AGN radiative feedback efficiency.
  \label{tab:params}}
  \centering
\begin{tabular}{l l l l l}

 \hline
     \scriptsize{\texttt{ID}} & \scriptsize{AGN Feedback} & \scriptsize{$\epsilon_{f,\mathrm{therm}}$} & \scriptsize{$\epsilon_{f,\mathrm{kin}}$} & \scriptsize{$\epsilon_{f,\mathrm{rt}}$} \\
 \hline
\scriptsize{\texttt{NoBH}} & \scriptsize{-} & \scriptsize{0} & \scriptsize{0} & \scriptsize{0} \\
\scriptsize{\texttt{NoAGN}} & \scriptsize{-} & \scriptsize{0} & \scriptsize{0} & \scriptsize{0} \\
\scriptsize{\texttt{Therm(Edd\_lim)}} & \scriptsize{Only Thermal} & \scriptsize{0.15} & \scriptsize{0} & \scriptsize{0}\\  
\scriptsize{\texttt{Therm}} & \scriptsize{Only Thermal (Edd-limited)} & \scriptsize{0.15} & \scriptsize{0}  & \scriptsize{0}\\
\scriptsize{\texttt{ThermKin}} & \scriptsize{Thermal + Kinetic} & \scriptsize{0.15} & \scriptsize{0.15} & \scriptsize{0} \\ 
\scriptsize{\texttt{ThermKinRad}} & \scriptsize{Thermal+Kinetic+Radiation} & \scriptsize{0.15} & \scriptsize{0.15} & \scriptsize{0.70} \\ 
\scriptsize{\texttt{ThermHEKin}} & \scriptsize{Thermal+High-energy Kin.} & \scriptsize{0.15} & \scriptsize{0.85} & \scriptsize{0} \\
\scriptsize{\texttt{ThermHEKinRad}} & \scriptsize{Thermal+High-energy Kin.+Radiation} & \scriptsize{0.15} & \scriptsize{0.85} & \scriptsize{0.70}  \\




  \hline
\end{tabular}
\end{table}




\textbf{Radiative Transfer:}
We use the \textsc{Ramses-RT} implementation by \citet{rosdahl_ramses-rt_2013} and \citet{2015MNRAS.449.4380R} for simulating the injection, propagation, and interaction of radiation with the multi-phase gas. 
%
Due to our relatively high spatial resolution of $\Delta x \approx 10\,\mathrm{pc}$, we expect well-resolved escape of ionizing radiation both from the model galaxy and within its ISM \citet{2014ApJ...788..121K}.
In its radiation hydrodynamics implementation, \textsc{RamsesRT} employs a first-order Godunov method with the $\mathrm{M1}$ closure \citep{Levermore1984,Dubroca&Feugeas1999} for the Eddington tensor. 
By using an explicit solver for the radiative transport, the advection time-step $\Delta t$, and consequently the CPU time, inversely scale with the speed of light $c$ as $\Delta t<\Delta x/3c$. This constraint mandates a time step significantly shorter than the hydrodynamic time step, which is limited by the maximum velocity of the gas ( $\sim1000\,\mathrm{kms^{-1}}$), rather than the speed of light. 
To mitigate this constraint, we adopt the reduced speed of light approximation \citep{2001NewA....6..437G}. 
We set the reduced speed of light at $0.2\,c$, with the radiation solver subcycling over the hydrodynamical time-step up to a maximum of $500$ steps. This adjustment has been found to be sufficient for modeling the propagation of ionization fronts through the ISM of galaxies \citep{rosdahl_ramses-rt_2013}.


In this work, using the configuration in \textsc{SPHINX} simulations \citep{rosdahl_lyc_2022}, we consider three photo-absorbing species of hydrogen and helium.
%
The radiation groups are divided into spectral bins as 
\begin{equation}
    \mathrm{Photon\, group}= 
    \begin{cases}
    \mathrm{HI} & 13.6\,\mathrm{eV}<\epsilon_{\mathrm{photon}}<24.59\,\mathrm{eV}\\
    \mathrm{HeI} & 24.59\,\mathrm{eV}<\epsilon_{\mathrm{photon}}<54.42\,\mathrm{eV}\\
    \mathrm{HeII} & \epsilon_{\mathrm{photon}}>54.42\,\mathrm{eV}.\\

    \end{cases}
\end{equation}

%

%
In our simulations, stellar particles and AGN are the sources of ionizing radiation. 
For the luminosity of each stellar particle we use the Binary Population and Spectral Synthesis (\textsc{bpassv2.0}) model \citep{2008MNRAS.384.1109E,2016MNRAS.456..485S} to radiate energy into its hosting cell with a spectral energy distribution (SED) according to particle mass, metallicity and age. The radiation from AGN is described in detail later in this section.

\textbf{Ideal MHD:} 
In the ideal MHD setup, \textsc{Ramses} employs a Constrained Transport method \citep{Teyssier2006,Fromang2006} to solve the equations that govern the evolution of the magnetic fields.
The induction equation is solved in a conserved integral form on the cell faces. This requires magnetic fields to be stored as six fields on the cell faces. This is unlike all the hydrodynamic quantities in the simulation, namely, densities, velocities, and energy components which are stored at the center of each gas cell.
The CT method ensures that the magnetic field has zero divergence down to numerical precision, preventing any unwanted modifications of conserved quantities \citep{2000JCoPh.161..605T} or the emergence of magneto-hydrodynamical artifacts \citep{2016MNRAS.455...51H}.

In the ideal MHD setup, assuming a highly conductive medium, the induction equation is solved with negligible diffusivity. Additionally, the distortion of the primordial field due to the velocity perturbations is only significant at sufficiently small scales and can be disregarded at galactic scales \citep[][]{1978PhDT........95W, 1980lssu.book.....P}. 
In the absence of non-ideal magnetic sources such as 
the Biermann battery, 
this leads to ${\partial\, (a^2\vec{B}_{\lambda})}/{\partial t}=0$ 
for the time evolution of the field.   
As the growth of compressional modes is suppressed before recombination, it solves as $\vec{B}_{\lambda}(\vec{x}, t) = \vec{B}_{\lambda}(\vec{x},t_{\mathrm{rec}})\, a^2(t_{\mathrm{rec}})/a^2(t)$, where
$\vec{B}_{\lambda}(\vec{x},t_{\mathrm{rec}})$ refers to the value of the magnetic field at recombination ($t=t_{\mathrm{rec}}$).
In this work, magnetic fields are modeled by an ab-initio $\vec{B}_{\lambda}$ that is seeded uniformly and aligned with the $z$-axis of the computational domain.
In the context of studying galaxy evolution and the growth of its central SMBH, this setup represents a primordial seed for magnetic fields with an initial comoving strength of $B_{\lambda}=10^{-21}\,\rm{G}$ coherent on large scales of $\lambda\sim1\,\rm{Mpc}$, consistent with the lower range of fields produced by the Biermann battery effect. 
In addition to primordial fields, in each supernovae (SNe) event, small-scale circular loops of magnetic fields are injected in the ISM, which are described in more detail later in this section.  


\textbf{Radiative cooling and heating processes:}
The hydrodynamical evolution of the gas is coupled to the local ionization via radiation pressure and the non-equilibrium hydrogen and helium thermochemistry, as described by \citet{rosdahl_ramses-rt_2013}.
%
%
In addition to primordial gas cooling, we account for metal-line cooling according to the gas metallicity.
Above temperatures of $10^4\,\mathrm{K}$ we interpolate the pre-calculated tables of  {\sc cloudy} \citep{Ferland1998}, assuming photoionization equilibrium with a redshift-evolving UV background. 
Below $10^4\,\mathrm{K}$ we follow fine structure metal cooling rates from \citet{Rosen1995}.
We model the process of reionization, based on the prediction from \citet{Haardt1996}, using a redshift-dependent UV background, which we initialize at redshift $z = 9$. 
Hydrogen self-shielding against the ionizing radiation is incorporated by 
suppressing the UV-background heating for gas densities above $n_{\mathrm{H}}=0.01\,\mathrm{cm}^{-3}$.
%
As the Pop III.1 star forms in metal-free gas, we employ a low-metallicity floor of $\left[ \mathrm{Fe}/\mathrm{H} \right] = -5$. 
Above this metallicity, fine-structure line cooling of atomic carbon and
oxygen is assumed to lead to gas fragmentation and the formation of low-mass stars \citep{Bromm2003}.

\textbf{Star formation:}
We model star formation employing a magneto-thermo-turbulent (MTT) star formation prescription, presented in more detail in the hydrodynamical version by \citet{Kimm2017} and \citet{Trebitsch2017}, and in its MHD version by \citet{Martin-Alvarez2020}. This comprehensive star formation model is particularly important for simulations of low-mass galaxies \citep{sanati_dwarf_2024}.
%
%
In summary, gas is allowed to transform into stellar particles only in cells that are at the highest level of refinement \citep{Rasera2006}, and where the local gravitational pull overcomes the combination of magnetic, thermal and turbulent support.
Then 
conversion of gas into stars in star forming cells
follows the Schmidt law \citep{Schmidt1959},
\begin{equation}
    {\dot\rho_{\star}} = \epsilon_{\mathrm{ff}}\frac{\rho_{\mathrm{gas}}}{t_{\mathrm{ff}}},
\end{equation}
where $\rho_{\mathrm{gas}}$ is gas density, $t_{\rm ff}$ is the local free-fall time, and $\epsilon_{\mathrm{ff}}$ is the efficiency of star formation per free-fall time.
In our formulation, $\epsilon_{\mathrm{ff}}$ is not a constant, but rather computed locally based on the gas properties of each region using the multi-scale model of \citep{2011ApJ...730...40P} as presented in \citet{Federrath2012}. 


%

\textbf{Stellar feedback:}
Each stellar particle in our simulations corresponds to a single stellar population that is characterized by an initial mass function (IMF).
The IMF is modeled as a probability distribution function following \citet{Kroupa2001} and normalized over the complete range of masses.
This allows each stellar particle to have a mass of 
$\sim2.4\times10^2 - 7\times10^3\,M_{\odot}$, which is then stochastically populated with stars within the mass interval of $\left[0.05 - 50\right]\,M_{\odot}$ during the initial $50\,\mathrm{Myr}$ of its formation. 
%
The number of exploding SNe for each particle is then calculated at each time step based on the lifetimes of the stars contained. 
%

We apply the mechanical supernova feedback model of \citet{2014ApJ...788..121K}. 
This method involves ejecting momentum in supernova explosions, which is determined by the physical characteristics of the gas being swept up, such as its density $n_{\mathrm{H}}$ and metallicity $Z$, 
\begin{equation}
    p_{\mathrm{SN}}(E,\, n_{\mathrm{H}},\, Z) \approx 3\times 10^{5}\,\mathrm{km}\,\mathrm{s}^{-1}\,\mathrm{M}_{\odot}\,E_{51}^{16/17}\,n_{\mathrm{H}}^{-2/17}\,f(Z),
\end{equation}
where the momentum input decreases with metallicity as $f(Z) =\mathrm{max}\left[Z/Z_{\odot}, 0.01\right]^{-0.14}$ and $E_{51}$ is the energy in the unit of $10^{51}\,\mathrm{erg}$.
%
This method 
ensures that the feedback from supernovae is accurately modeled at all blast wave stages, from 
the initial free expansion to the final momentum-conserving snowplow phase.
Along with the momentum, energy is deposited into the neighboring cells. 
The specific energy of each SNe has a value 
$\varepsilon_\text{SNe} = E_\text{SNe} / M_\text{SNe}$, where 
 $E_\text{SNe} = 10^{51} \mathrm{erg}$ and $M_\text{SNe} =10\, \mathrm{M}_{\odot}$.
Each SNe also returns a fraction of stellar mass back to the ISM.
We use $\eta_\text{SNe} = 0.2$ for fraction of $M_\text{SNe}$ returned as gas mass, and $\eta_\text{metal} = 0.075$ for the newly synthesized metals. 

We use the magnetized stellar feedback prescription of \citet{martin-alvarez2021}. In this approach, three pairs of magnetic field loops are injected into the cells surrounding a stellar particle undergoing a supernova. The magnetic energy released is adjusted to be approximately $1\%$ of total supernova energy, $E_\text{SNe}$. These magnetized loops, initially spanning $\sim 10\,\mathrm{pc}$, generate a magnetic field with a strength of $\gtrsim10\,\mathrm{\mu G}$ in each supernova event, consistent with observed high magnetization of supernova remnants \citep{parizot2006} \citep[see also][]{2013pss5.book..641B,2017ApJ...843..113B,vazza2017}.

In summary, our stellar feedback model incorporates radiation, as discussed earlier in this section, alongside magnetic, kinetic, and thermal energy inputs. 
We note that our feedback prescription does not include stellar winds and cosmic rays, both of which, along with SNe feedback, could potentially reduce the final stellar mass of our galaxy model. However, dwarf galaxy simulations \citep{martin-alvarez_pandora_2022,2025arXiv250303813R} show that radiation feedback has a greater impact than these other feedback mechanisms.


\textbf{The SMBHs formation:}
We designate the first collapsing dark matter minihalo as the site of formation for the Pop III.1 star. This star is formed once our MTT star formation criterion described above is fulfilled inside this minihalo.
The Pop III.1 star is modeled as a single stellar particle with a mass of $10^5\,M_{\odot}$. This mass is extracted from the baryonic gas mass in this minihalo, and constitutes a large proportion of its total baryonic content.
During its lifetimes, the Pop III.1 star has a phase of intense H-ionizing photon emission, with rates of $\sim 10^{53}\,\rm{s}^{-1}$ sustained for $\sim 10^7\,\rm{yr}$. 
At the end of its lifetime, a Pop III.1 star is expected to undergo core collapse to form a SMBH with high mass conversion efficiency. 
To model this in the simulations, we turn off the radiation from the star and replace it with a SMBH sink particle with the same mass of $10^5\,M_{\odot}$.
For seeding the SMBH in our simulations we employ our modified version of the implementation by \citet{Dubois:2010}. 


%



\textbf{SMBHs Accretion:}
Once the sink particle is formed, it is divided into multiple cloud particles. 
Using $2109$ clouds distributed every $0.5\,\Delta x$ around a sink particle, we tile a sphere with a radius $r_{\mathrm{cloud}}=4\,\Delta x$, effectively extending to the comoving radius of $\sim$$40\,\mathrm{pc}$ around the BH at our resolution.
Cloud particles effectively sample local cells and provide averaged gas properties, while the sink particle is moving across the grid. These averaged quantities are then used to compute the BH accretion.
We use the Bondi-Hoyle-Lyttleton \citep[hereafter BHL,][]{(Hoyle&Lyttleton1939, Bondi&Hoyle1944, Edgar2004} formula to calculate the BH accretion rate:
\begin{equation}
    {\dot M}_{\mathrm{BH}} = 4\pi\, {\rho_{\infty}}\,v_{\mathrm{BHL}}\,r_{\mathrm{BHL}}^2, 
\end{equation}
where ${\rho_{\infty}}$ is the gas density sufficiently far from the gravitational effects of the sink particle,
$v_{\mathrm{BHL}}$ is the Bondi-Hoyle-Lyttleton velocity give by
\begin{equation}
v_{\mathrm{BHL}} = \sqrt{\overline{c}^2_s+\overline{v_{\mathrm{rel}}}^2},
\end{equation}
where $\overline{c}_{s}$ is the average sound speed and $\overline{v}_{\mathrm{rel}}$ is the average of gas velocity relative to the black hole (these are computed using the volume-weighted contribution of cloud particles), and $r_{\mathrm{BHL}}$ is the Bondi radius given by
\begin{equation}
r_{\mathrm{BHL}}= {G M_{\mathrm{BH}}}/{v_{\mathrm{BHL}}^2}.
\end{equation}
%

%
$G$ is the gravitational constant and $M_{\mathrm{BH}}$ is the mass of the black hole. 
To relate the gas density at infinity to the average gas density $\overline{\rho}$, we use a Gaussian kernel first introduced in tabulated form by \citet{Krumholz2004}.  
\begin{eqnarray}
w_{\mathrm{cl}} = \mathrm{exp}\,\left(-\frac{r_\mathrm{cl}^2}{r_{s}^2}\right),
\end{eqnarray}
%
where $\rho_{\infty} = \overline{\rho}/w_{\mathrm{cl}}$. This way the weighting of cloud particles depends on the distance $r_\mathrm{cl}$ between the gas cell containing each cloud particle and the sink particle. The scale radius $r_s$ is derived from the minimum cell size and the Bondi radius,
\begin{equation}
    r_s= 
    \begin{cases}
    \Delta x/4 & r_{\mathrm{BHL}}<\Delta x /4\\
    r_{\mathrm{BHL}} & \Delta x /4 \leq r_{\mathrm{BHL}} \leq 2\Delta x\\ 
    2\Delta x & r_{\mathrm{BHL}} > 2\Delta x 

    \end{cases}
\end{equation}

The accretion in the \texttt{Edd\_lim} model (see Tab.~\ref{tab:params}) is limited by the Eddington rate, $\dot{M}_{\mathrm{Edd}}$, which depends solely on the BH mass. 
In the other models examined in this work, the accretion rate is not numerically capped by the Eddington limit; instead gas accretion onto the BH is regulated by AGN feedback.
%
%
Although using an extra artificial boost for the accretion rate is a common practice in lower resolution cosmological simulations, the spatial resolution in this work is sufficient to capture the Bondi radius
and does not require an artificially boosted accretion.

Once the accretion rate to the sink particle ${\dot M}_{\mathrm{BH}}(\rho_{\infty},\overline{v}_{\mathrm{rel}},\overline{c}_{s})$ is obtained, at each time step $\Delta t$, the gas mass is transferred from the cells containing the clouds to the sink particle in a volume-weighted scheme,
\begin{eqnarray}
    \Delta m_\mathrm{cl}= \frac{\rho_{\mathrm{cl}}\times V_{\mathrm{cl}}}{\overline{\rho}\times\overline{V}}\, {\dot M}_{\mathrm{BH}}\, \Delta t.
\end{eqnarray}
%
This is carried out by iterating over all cloud particles in the grid. 



\textbf{AGN feedback quasar and radio modes:}\label{AGNfb}
%
Similar to the dual-mode implementation of Horizon-AGN \citet{dubois:2012}, AGN feedback depends on whether the accretion rate is comparable to the Eddington limit or below one percent of $\dot{M}_{\mathrm{Edd}}$. 
In the former case, the BH is in the quasar mode, where a large fraction of the mass flow to the accretion disc is efficiently radiated away. The energy release in this mode is primarily dominated by thermal and radiative energy, with a smaller (but non-zero) contribution from kinetic energy which drive outflows and winds.
In the latter case, the accretion flow to the BH is radiatively inefficient and the BH is in the radio mode. Thus the energy is released only in the form of kinetic energy.
We do not include an explicit duty cycle.
In both modes (labeled by $x$), the fraction $\epsilon_{f,x}$ of the radiated energy $L_{\mathrm{AGN}}$ from the accreting gas that is deposited back to the surrounding medium as thermal or kinetic feedback is given by
 \begin{eqnarray}
\dot{E}_{\mathrm{AGN}}=\epsilon_{f,x}\,L_{\mathrm{AGN}}=\epsilon_{f,x}\epsilon_r\,{\dot M_{\mathrm{BH}}}\,c^2, 
\label{eq:LAGN}
\end{eqnarray}
%
where $\epsilon_r=0.1$ is the radiative efficiency of the accretion disc. 
$\epsilon_{f,x}$ is mode dependent and its value in each model is given in Tab.~\ref{tab:params}.
%
%
For the kinetic feedback, we set $\epsilon_{f,\mathrm{kin}}=0.15$ and $0.85$ to represent both the weak and strong radio feedback models.  
Along with the kinetic energy, momentum is deposited with a velocity of $u_{\mathrm{J}}=10^4\,\mathrm{km\,s^{-1}}$, weighted by the distance of the clouds to the sink particle, within the cloud radius.   
This mechanical feedback is initially deposited isotropically, without a defined cone opening. However, the outflows naturally become collimated by the interstellar medium of the galaxy and develop into a bipolar structure.

\textbf{AGN radiation:}\label{AGNrad}
In addition to the thermal and kinetic energy ejection described in the previous section, we release radiation energy from the BHs to account for the contribution of AGN to the ionizing radiation field. For this, we applied the method presented in \citet{Bieri:2017}. We release radiation at each fine timestep, and the amount of radiation released in each frequency bin is given by the luminosity of the quasar in each band.    
%
%
From the broad-band SED adopted in the unobscured spectrum of \citet{Sazonov:(2004)}, we calculate the corresponding fraction of the energy distribution, which is then multiplied by the quasar luminosity to yield the corresponding group energy for each photon group,
\begin{eqnarray}
L_{\mathrm{rad}}=f_{\mathrm{UV}}\epsilon_r\,{\dot M_{\mathrm{BH}}}\,c^2, 
\end{eqnarray}

%

with $f_{\mathrm{UV}} = f_{\mathrm{UV},\mathrm{HI}} + f_{\mathrm{UV},\mathrm{HeI}} + f_{\mathrm{UV},\mathrm{HeII}}$, where, 
for each photon group, the cross-sections are luminosity-weighted averages over the energy interval, as described in \citet{rosdahl_ramses-rt_2013}.
%
%
In these simulations we do not directly model the infrared emission from the dust, and therefore focus on the AGN ionizing luminosity in the UV bands. Instead, the release of thermal energy, described in the previous section, represents the AGN luminosity in these two absent bands, since they are not energetic enough to ionize the hydrogen or helium.
%



\textbf{BH dynamics:}
In this work, the BH is neither pinned to the center of mass of its host galaxy \citep{Sijacki:2007} nor artificially directed towards the center of its host dark matter halo \citep{costa_feedback_2014}. 
Instead, we allow it to move freely across the AMR grid, with its trajectory determined by the gravitational pull from the surrounding matter. 
The accretion of gas onto the BH transfers additional momentum to the sink particle. Similarly, the relative velocity between the gas and the sink particle generates dynamical friction, leading to further momentum transfer between the BH and the gas.

In this scheme, the gas attracted towards the sink particle by accretion, generates a local over density behind the moving BH. This additional gravitational pull from this over-dense gaseous region acts as a drag force  
increasing the likelihood of the sink particle staying in the gravitational well of the dense local gas.
The resolution in this work is sufficient to resolve this drag force, allowing the simulations to avoid the common issue of BH ejection from the galactic disk, a problem frequently encountered in galaxy simulations \citep{Sijacki:2007, Volonteri:2016, Biernacki:2017,2021MNRAS.500.4639B}.
%


\textbf{Initial conditions and model resolution:}
The initial conditions are generated using the \textsc{MUSIC} code \citep{2011MNRAS.415.2101H},
and the cosmology of \citet{2016A&A...594A..13P} with $\Omega_\Lambda = 0.6911$,  
$\Omega_m = 0.3089$, $\Omega_b = 0.0486$, and $h = 0.678$.
All simulations are started at redshift $z = 100$, ensuring that the rms variance of the initial density field, lies between 0.1 and 0.2 \citep{2009ApJ...698..266K,onorbe2015}. 
%
%
%
Using the zoom-in technique, for the halo studied in this work we refine a 3D ellipsoid of size about $0.85\,\mathrm{cMpc}$ 
across, positioned in the center of the cubic simulation box with $L_{\mathrm{Box}}=7.55\,\mathrm{cMpc}$ per side.
The size of the ellipsoid is determined such that it encompasses all particles that eventually reside within the target halo by redshift $z=0$. 
In this refined region, we achieve dark matter mass resolution of 
$m_{\mathrm{DM}}\simeq10^{4}\,\mathrm{M}_{\odot}$.
We gradually degrade the resolution, from a minimum level of 10 within the zoom region, to 9 outside \footnote{One resolution level $l$ corresponds to $N=(2^l)^3$ particles in the full cosmological box. The particle mass is thus decreased by a factor of eight between two levels.}.

Initially, the domain is discretized with a uniform grid of $1024^3$ cells. This resolution is preserved within the zoom region, while the grid is de-refined to level $9$ elsewhere. 
Throughout the course of the simulation, the adaptive refinement criteria come into play to effectively resolve dense and Jeans-unstable regions.
When the total dark matter and gas mass within a grid cell exceeds $8\,m_{\mathrm{DM}}$, or when the size of the grid cell surpasses $4$ local Jeans length, a parent grid cell is split into $8$ equal child cells. 
This process follows the octree structure of \textsc{Ramses}, where
the size of cell $i$ is determined by the refinement level of cell $l_i$ according to $\Delta x_{i} = 1/2^{l_i}\,L_{\mathrm{Box}}$. 
In our simulations, with a maximum refinement level of $19$, the initial grid undergoes adaptive refinement to achieve a minimum cell width of approximately $14\,\mathrm{pc}$.


\section{Results}\label{sec:res}

\begin{figure*}
    \centering
    \includegraphics[width=0.72\textwidth]{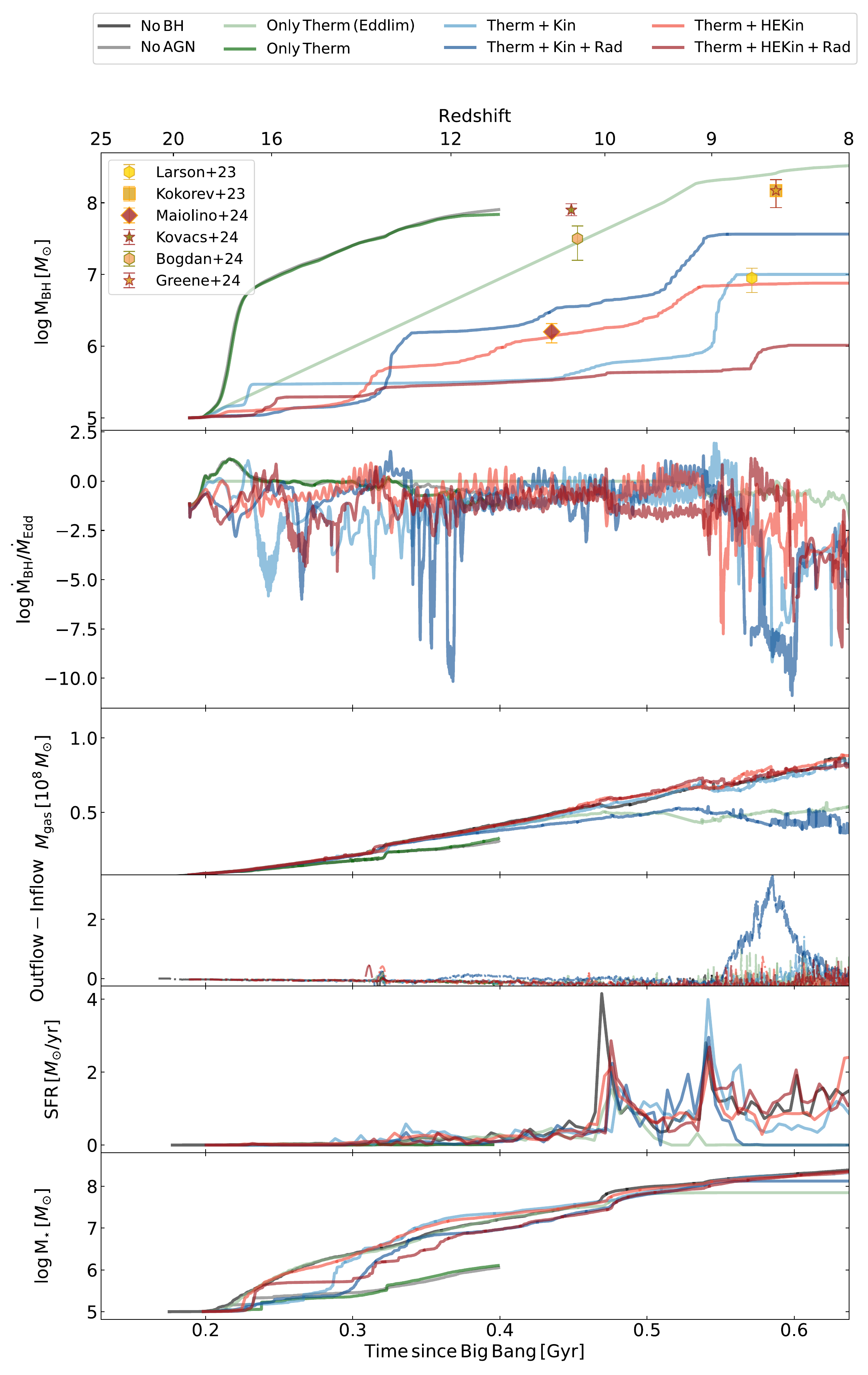}

    \caption{The effect of AGN feedback components on the time evolution of BH-galaxy properties. From top to bottom, the rows display BH mass, Eddington fraction, gas mass, gas net flow, star formation rate, and stellar mass, within $0.15\,R_{\mathrm{vir}}$.
    The BH masses are compared to the progenitor mass of observed quasars at redshifts $z>8$ \citep{Larson:2023, Kokorev:2023,  2024A&A...691A.145M, 2024ApJ...965L..21K,2024NatAs...8..126B,2024ApJ...964...39G}. 
    %
    %
    The short super-Eddington accretion episodes, reaching up to $86$ times $\lambda_{\mathrm{Edd}}$, 
    are followed by a sharp decline, caused by AGN feedback.
    The peak in the net gas flow into the galaxy in the \texttt{ThermKinRad} model shows the role of 
    radiation besides kinetic winds
    in launching strong AGN-driven outflows.  
    reduced gas mass and stellar mass in the \texttt{NoAGN} model shows the impact of BH accretion on quenching star formation at redshifts $z>15$.
    } 
    \label{fig:BHStGs_acc}
\end{figure*}

\subsection{The role of AGN feedback in BH-galaxy co-evolution}\label{sec:smbh}

\begin{figure}
    \centering
    \includegraphics[width=0.48\textwidth]{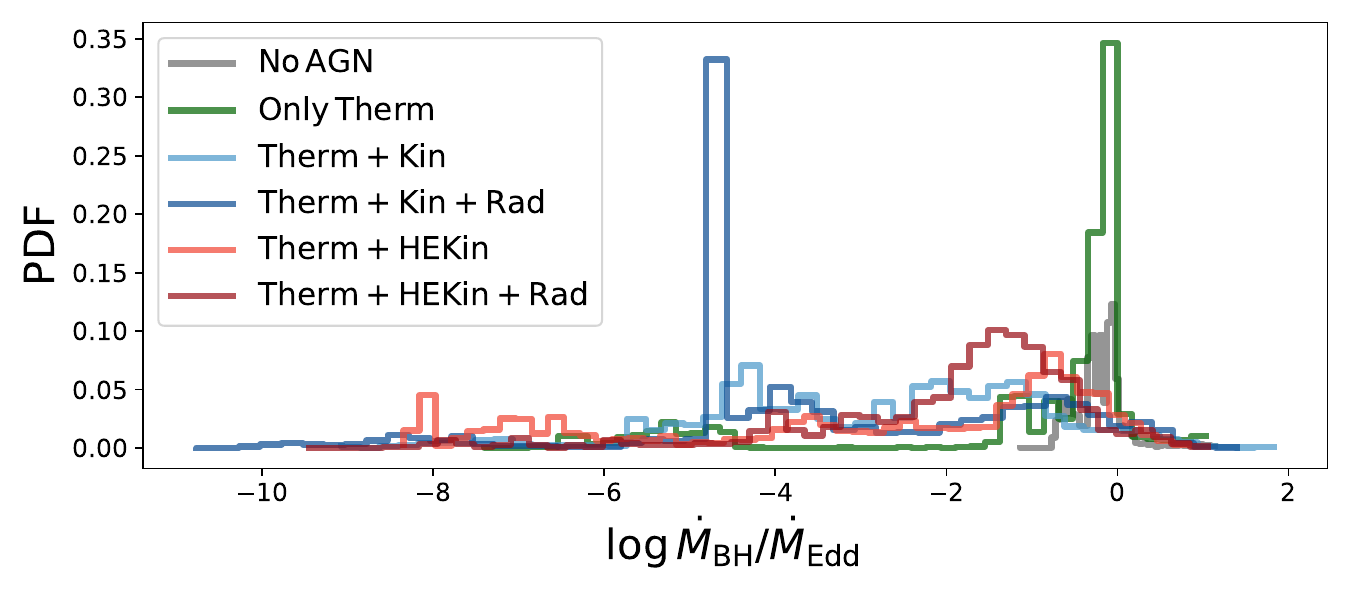}
    

    \caption{
    Probability distribution function (PDF) of the Eddington fraction, $\lambda_{\mathrm{Edd}}$.
    The accretion rate vary depending on the AGN feedback model. The BH in all models primarily grows through accretion rates that are below and comparable to the Eddington limit. Only a small fraction of the BH mass grows through super-Eddington accretion rates with the maximum of $\lambda_{\mathrm{Edd}}= 86$.} 
    \label{fig:fEddhist}
\end{figure}

The results for the BH accretion rate and final mass, along with some of the host galaxy properties, are presented in Fig.~\ref{fig:BHStGs_acc}.  
All the calculations are computed for the \textit{galactic region}. The physical radius of this region is approximately $6\,\mathrm{kpc}$ at redshift $z \sim 8$ which corresponds to $0.15$ times the virial radius of the host dark matter halo, $R_{\mathrm{vir}}$.
%
The black line in the stellar and gas mass panels represents the fiducial \texttt{NoBH} model. In the gray model the BH is formed but its AGN feedback is set to zero.
In the \texttt{Edd\_lim} model, in light green, AGN activity is constrained by limiting the accretion to the Eddington rate, with feedback released solely as thermal energy. 
In the other models, accretion is self-regulated by various feedback mechanisms. 
The light blue and orange lines represent weak and strong kinetic feedback, respectively, where the AGN ejects mechanical energy with efficiencies of $\epsilon_{f,\mathrm{Kin}}=0.15$ and $0.85$ 
in both radio and quasar modes (the feedback efficiency parameter is explained in Eq.~\ref{eq:LAGN}). 
The two most comprehensive models, shown in dark blue and red, include AGN radiation which complements the weak and strong radio feedback, with radiation included only in the quasar mode.

The top panel shows the evolution of the BH mass in different models compared to progenitor mass of six observed quasars at redshifts higher than $z=8$ \citep{Larson:2023, Kokorev:2023,  2024A&A...691A.145M, 2024ApJ...965L..21K,2024NatAs...8..126B,2024ApJ...964...39G}.
%
In the \texttt{NoAGN} model, the SNe feedback alone is unable to regulate the BH accretion. 
This results in a steep increase in BH mass, from an initial seed of $10^5$ to $5\times10^7$ in just $\sim100\,\mathrm{Myr}$. 
%
Since the exact impact of AGN feedback on BH accretion remains elusive, this unrealistic \texttt{NoAGN} model serves as a reference to see the impact of different feedback components. 
The similarity in BH growth between the \texttt{Therm} and \texttt{NoAGN} models show the inefficiency of AGN feedback when modeled only by releasing thermal energy due to its little effect on the gas surrounding the BH. 
The short time-stepping required by the high spatial resolution of our simulations reduces the artificial buildup of feedback energy before it is deposited onto the grid. This further reduce the effectiveness of thermal feedback, as the energy can be rapidly radiated away. 
We stop these two simulations at redshift $z\sim11$ due to the high computational cost. 
When the BH is not allowed to accrete above the Eddington limit, its mass grows steadily, drawing from an extensive gas reservoir, and maintaining a constant growth rate at high redshifts. By redshift $z\sim 9$, the BH reaches its final mass of $\simeq5\times 10^8\,M_{\odot}$. 
At this stage, the SMBHs has grown so massive that its mass is comparable to the total baryonic mass of the galaxy, suggesting it has already accreted nearly all of the available gas. Its growth slows not due to external regulation, but simply because little to no gas remains to be accreted.

In the other four models, the SMBHs accretion is self-regulated by AGN feedback, and the BH reaches a final mass between $\sim10^6$ and $5\times10^7\,M_{\odot}$ at redshift $z=8$. 
Amongst these four, the \texttt{ThermKinRad} model, which includes AGN radiation and kinetic winds, leads to the highest SMBHs final mass.
Although the final mass in this model is lower than in the \texttt{Edd\_lim} case, the resulting $M_{\mathrm{BH}}/{M_\star}$ ratio remains within the range  $0.01$ and $1.0$, consistent with observed scaling relations at high redshifts (further discussed below).
%
The BH has the lowest final mass in the \texttt{ThermHKinRad} model, where the same feedback components are present, but with the difference that momentum is released in 
the strong radio mode. As a result, 
kinetic winds are approximately six times stronger ($\epsilon_{f,\mathrm{Kin}}=0.85\,L_{\mathrm{AGN}}$) than in the \texttt{ThermKin} model.
%
The final BH masses 
of the two models without radiation, 
\texttt{ThermKin} and \texttt{ThermHKin}, lie between those of their counterparts with radiation \texttt{ThermKinRad} and \texttt{ThermHKinRad}, respectively. 
Variations in feedback strength, particularly the inclusion of radiation in the quasar mode, also affect star formation in the host galaxy. A reduced star formation rate, as in the \texttt{ThermKinRad} model, leaves more gas available for BH accretion. In this way, AGN radiation can enhance BH growth by suppressing star formation.
However, if the AGN also has powerful kinetic feedback in the radio mode, in addition to radiative feedback in the quasar mode, as in the \texttt{ThermHKinRad} model, the feedback becomes excessively strong, thereby preventing gas accretion onto the BH.
Nonetheless, one could argue that the overall BH growth across these four feedback-regulated models is broadly similar, with variations in final BH mass comparable to those arising from stochastic effects in galaxy evolution simulations.

The second panel shows the Eddington ratio, $\lambda_{\mathrm{Edd}}$, which is the ratio of the Bondi-Hoyle accretion rate, as used in the simulation, to the Eddington limit which varies proportionally with the BH mass. The light green line, which maintains $\lambda_{\mathrm{Edd}}=1$ for most of the BH evolution, represents the \texttt{Edd\_lim} model. 
In all models, the accretion rate right after BH formation is sub-Eddington for a few $\mathrm{Myrs}$.
Fig.\ref{fig:BH_acc} 
zooms in on the first $80\,\mathrm{Myr}$ to better highlight this initial accretion phase. 
It compares the accretion rate onto a $10^5\,M_\odot$ BH formed with and without a Pop III.1 star. 
Over the $\simeq150\,\mathrm{Myr}$ (between $200$ and $350\,\mathrm{Myr}$ after the Big Bang), BH grows to between two and ten times its initial mass in the AGN-regulated models, \texttt{ThermKin}, \texttt{ThermKinRT}, \texttt{ThermHKin}, and \texttt{ThermHKinRT}, and exceeds a hundredfold increase in the absence of AGN feedback.
%
The BH mass growth is later interrupted by a merger event between the BH host halo and a smaller neighboring halo occurring between redshifts $z = 15$ and $z = 11$. The accretion rate increases again at redshift $z \sim 9.5$, following this merger event. We discuss the BH and galaxy properties during and after the merger event in detail in \S.~\ref{sec:merger}.
%
%
%
The time evolution of $\lambda_{\mathrm{Edd}}$ shows short super-Eddington accretion episodes, reaching up to $86$ times $\lambda_{\mathrm{Edd}}$. These high accretion phases are followed by a sharp decline, caused by AGN feedback triggered during the peak.
For better understanding these episodes of low accretion, we look at the gas mass available for BH to accrete as well as the trajectory of the BH within its host galaxy.

Panel three in Fig.~\ref{fig:BHStGs_acc} shows the total gas mass within the \textit{galactic region}, which corresponds to $0.15\,R_{\mathrm{vir}}$. 
%
Panel four shows the gas net flow, calculated as outflowing gas minus inflowing gas into the galaxy, in the same region. 
%
The increase in the gas net flow into the galaxy at redshift $z<9$ following a merger event leads to a regulatory cycle of: i) high BH accretion rates; ii) triggering of strong AGN feedback; iii) AGN-driven high velocity outflow; iv) depletion of gas from the galactic region; and v) a subsequent drop in BH accretion. This cycle and the resulting outflows are discussed in detail in Sec.~\ref{sec:sEdd} and Sec.~\ref{sec:outflows}, respectively.
This cycle is particularly pronounced in the \texttt{ThermKinRad} model, with a net flow of $\sim3\times10^8\,M_{\odot}/\mathrm{yr}$.
In this model, super-Eddington accretion drives strong outflows and gas depletion, resulting in a significant follow-up decrease in the accretion rate of the SMBH. However, it is important to also consider the role of SNe feedback in driving gas out of the central region of the galaxy, as this phase of high accretion, due to a merger, coincides with a peak in the star formation rate.

Panel six in Fig.~\ref{fig:BHStGs_acc} shows the star formation rate. 
All feedback models have a star formation history similar to the fiducial \texttt{NoBH} model. The star formation in this latter model is not influenced by BH accretion or AGN feedback. Thus, the peak amplitude at $z\sim10$, which marks the first star formation burst, is higher in this model compared to the other models that include BH formation. 
%
%
The main star formation bursts occur between redshift $z=11$ and $9.5$, and it coincides with the merger of the BH host halo and a satellite galaxy initially in a distance of $\sim10\,\mathrm{kpc}$ at $z=11$. 
The mass brought in by the satellite is $\sim8\%$ of the mass of the main galaxy in which the BH resides.
Although the merger leads to an increase in the star formation rate, the significant outflow of gas at $z<9$, specially in the \texttt{ThermKinRad} model, 
is more closely linked to the AGN feedback than to stellar feedback. 
This is evidenced by two measurements. Firstly, the peak in star formation at $z\sim10$ does not produce comparable outflows, as is particularly evident in the \texttt{NoBH} model. 
Secondly, during the outflow episode in the \texttt{ThermKinRad} model at $z<9$, the gas velocities reach $\sim3500\,\mathrm{km\,s^{-1}}$, far exceeding the typical velocities of  a few hundred $\mathrm{km\,s^{-1}}$ generated by SNe feedback \citep[see e.g.,][]{costa_quenching_2018, nelson_first_2019}. 
We discuss the outflow properties in detail in Sec.~\ref{sec:outflows}. 


The bottom panel of Fig.~\ref{fig:BHStGs_acc} shows the total stellar mass 
within the \textit{galactic region}, defined as $0.15\,R_{\mathrm{vir}}$. This region also encompasses the neighboring minihalos that later merge with the host galaxy.  
%
The final stellar mass in the \texttt{ThermKin}, \texttt{ThermHKin}, and \texttt{ThermHKinRT} are very similar to that in the fiducial \texttt{NoBH} model, with only minor differences in their evolution. 
In contrast, the final stellar mass in the \texttt{Edd\_lim} model is $\sim5$ times lower than the \texttt{NoBH} model, as the BH has accreted nearly all  of the available gas, leaving little fuel for star formation. 
An even more extreme trend is seen in the \texttt{NoAGN} and \texttt{Therm} models, where the BH mass growth extremely rapidly. 
%
%
The only difference between the \texttt{Edd\_lim} model and the \texttt{Therm} model is the imposition of an Eddington limit on the BH accretion rate. This Eddington limit leads the \texttt{Edd\_lim} model to feature $\sim1.6$ higher gas mass, and $20$ times higher stellar mass than the \texttt{Therm} model. This Eddington limit prevents rapid gas depletion characteristic of short bursts of super-Eddington accretion, observed in the \texttt{Therm} model at redshifts $z>15$.
This shows the impact of BH accretion, even in the absence of efficient AGN feedback, on quenching star formation at high redshifts, primarily because efficient accretion depletes gas from the high-density regions within the galaxy, leaving little available for star formation.
Similarly in the \texttt{ThermKinRad} model, which incorporates AGN radiation, the stellar mass is further reduced at lower redshifts.
This highlights the great impact of radiation, compared to other feedback components, in suppressing star formation when BH growth is sufficient, specially in low-mass galaxies. 
We discuss the importance of radiation pressure in AGN feedback in more detail in Sec.~\ref{sec:outflows}.


\begin{figure*}
    \centering
    \includegraphics[width=0.96\textwidth]{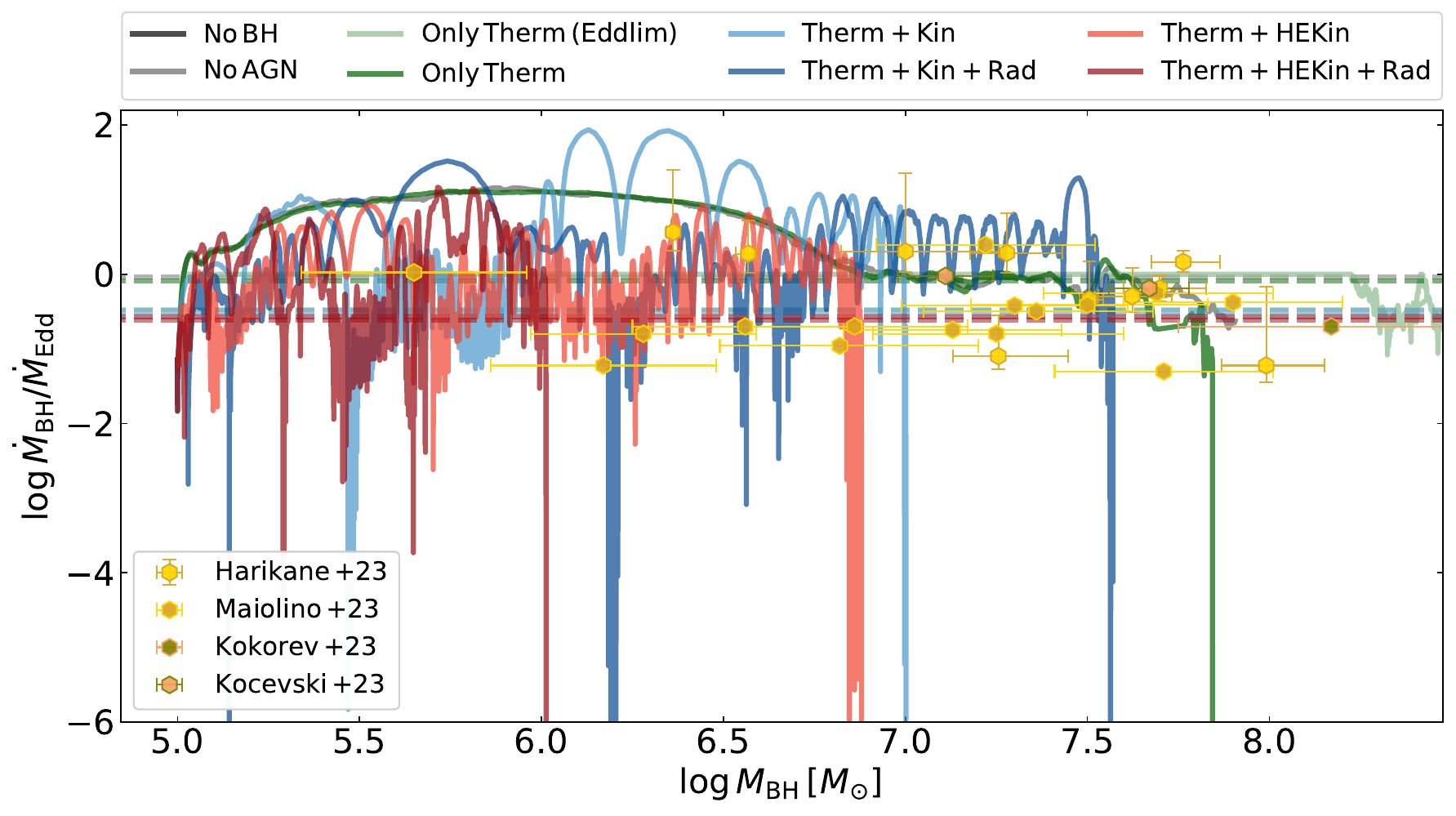}

    \caption{
    The Eddington fraction versus BH mass compared to the observed JDEEP sample in yellow hexagons.
    $\dot{M}_{\mathrm{BH}}$ represents the black hole accretion rate following the Bondi-Hoyle model, while the $\dot{M}_{\mathrm{Edd}}$ represents the accretion to the black hole if it follows the Eddington limit. 
    The green line shows \texttt{Edd\_lim} model where the AGN activity is Eddington limited. In other models the accretion is self-regulated by different feedback models.
    In the weakest AGN model \texttt{Therm} in dark green, where the feedback is limited to releasing thermal energy in the quasar mode without ejecting mechanical feedback or emitting radiation, 
    most of the BH mass is accreted in the initial super-Eddington phase in the first $\sim60\,\mathrm{Myr}$. The accretion rate in other models, regulated by a more comprehensive AGN feedback, leads to a more gradual build up the BH mass.  
    }
    \label{fig:fEddMBH}
\end{figure*}

\begin{figure}
    \centering
    \includegraphics[width=0.48\textwidth]{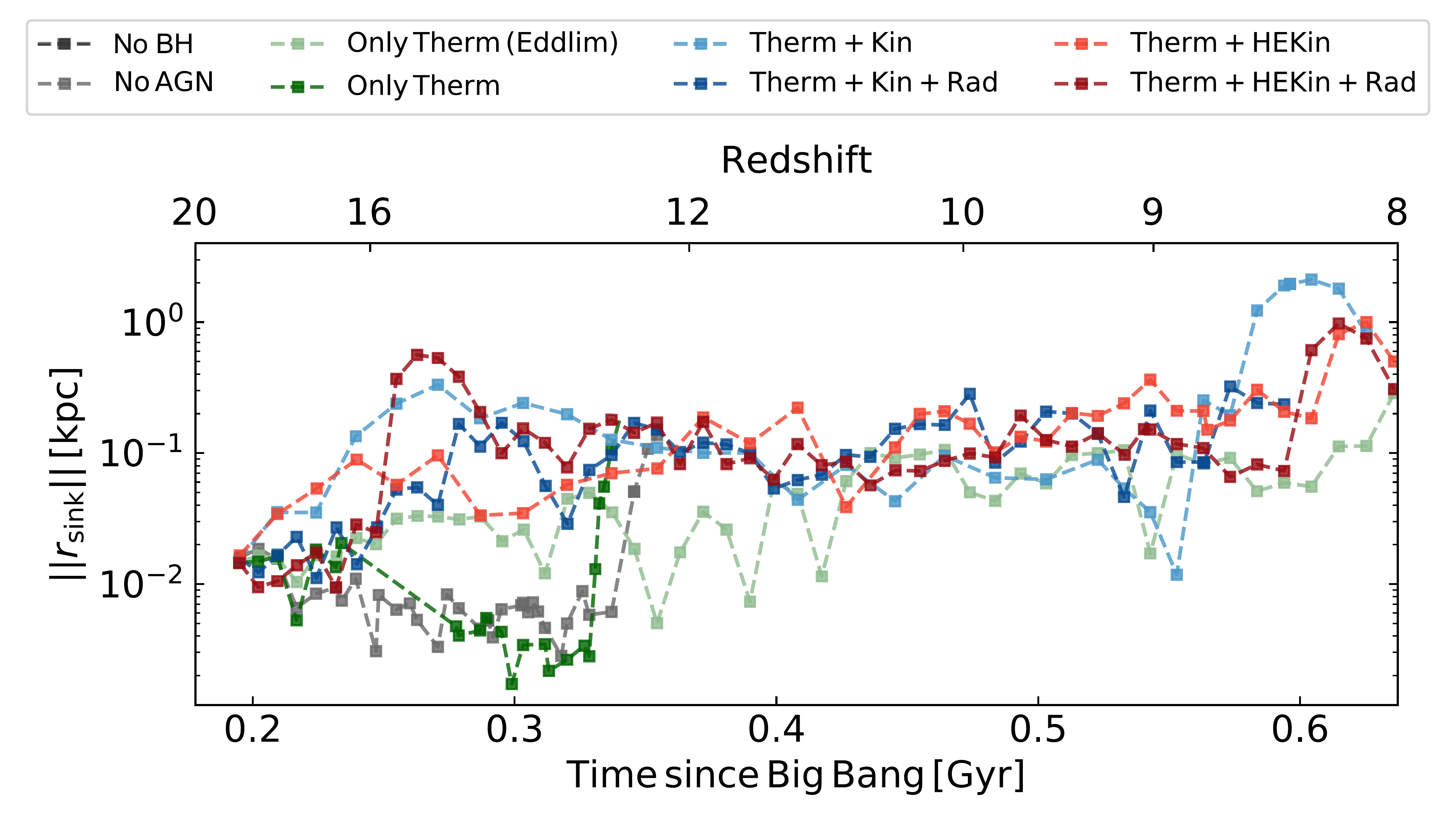}
    

    \caption{
    Time evolution of the BH relative displacement with respect to the center of the galaxy, and its variation depending on the AGN feedback model. For the BH to grow efficiently, its trajectory must intersect with dense gas clouds, primarily concentrated in the central region of the galaxy.
    On average across its evolution, the BH remains at a distance of a few hundred $\mathrm{pc}$ from the galaxy center.
    }
    \label{fig:sinkpos}
\end{figure}

\subsection{Feedback-driven cycles of super- and sub-Eddington accretion}\label{sec:sEdd}
%
As discussed in Sec.~\ref{sec:smbh}, BH growth follows a high accretion - strong feedback - low accretion cycle rather than maintaining a sustained high accretion rate.
To illustrate this cycle more quantitatively, Fig.~\ref{fig:fEddhist} shows the probability distribution function (PDF) of $\lambda_{\mathrm{Edd}}$ for the feedback-regulated models, compared with the \texttt{NoAGN} and \texttt{Therm} models. 
The shape and the location of the peak in the PDFs vary depending on the AGN feedback model.
Although all models contain episodes of super-Eddington accretion rates ($\lambda_{\mathrm{Edd}}> 1$), in the self-regulated feedback models the BH primarily grows at rates below yet close to the Eddington limit.
In contrast, the \texttt{NoAGN} and \texttt{Therm} models have PDF peaks at super-Eddington values, showing that when the AGN emits weak feedback, there is little to no prevention of BH accretion, allowing it to grow rapidly.   
%
%
%
In all AGN feedback-regulated models, the PDFs peak near $\lambda_{\mathrm{Edd}}\sim 1$ which shows accretion predominantly occurs at rates comparable to the Eddington limit.
Additionally, the presence of lower $\lambda_{\mathrm{Edd}}$ values, $\lesssim10^{-2}\,M_{\odot}\,\mathrm{yr}^{-1}$, which is absent in the \texttt{NoAGN} model, shows the impact of AGN feedback in consistently suppressing accretion rates.  
The distinct shape of the distribution in AGN-regulated models, compared to \texttt{NoAGN} and \texttt{Therm}, further shows the role of AGN feedback in maintaining moderate accretion rates and preventing runaway growth.
While all feedback-regulated models allow only a small fraction of the BH mass to grow through super-Eddington rates, their accretion reach super-Eddington values with maxima of $\lambda_{\mathrm{Edd}}\sim 86,\,33,\,8.5$, and $15$ in the \texttt{ThermKin}, \texttt{ThermKinRad}, \texttt{ThermHKin}, and \texttt{ThermHKinRad} models, respectively. 
%



We also compare the accretion rates in our models to the JDEEP sample of high-redshift SMBHs \citep{Harikane:2023b, maiolino_jades_2023, Kokorev:2023, 2023ApJ...954L...4K}.
In Fig.~\ref{fig:fEddMBH} we show $\dot{M}_{\mathrm{BH}}/\dot{M}_{\mathrm{Edd}}$ as a function of BH mass for different feedback models. 
This plot illustrates the relative accretion rate at each stage of the BH growth, highlighting how the accretion rate changes as the BH gains mass throughout its evolution.
Dashed lines show the average accretion rate, $\overline{\lambda}_{\mathrm{Edd}}$, across BH mass in each model. 
%
%
Comparison with observational data shows that our feedback-regulated models successfully reproduce BH masses in the range of $\sim$$10^5$ to $5\times10^7\,M_{\odot}$ with average accretion rates comparable and below the Eddington limit. 
In the feedback-regulated models, 
\texttt{ThermKin}, \texttt{ThermKinRad}, \texttt{ThermHKin} and \texttt{ThermHkinRad},
$\overline{\lambda}_{\mathrm{Edd}}$$\sim 0.32,\,0.27,\,0.27$, and $0.24$, respectively.
In these models, the accretion rate fluctuates around the Eddington limit, with variations comparable to those arising from stochastic effects. 
These fluctuations around ${\lambda}_{\mathrm{Edd}}=1$ further emphasize the role of AGN feedback, alongside SNe feedback, in regulating the accretion rate and generating a cycle of high accretion followed by a strong feedback. Compared to the \texttt{Edd\_lim} model, this cycle leads to a more gradual increase in BH mass. 
In contrast, the \texttt{NoAGN} model, with $\overline{\lambda}_{\mathrm{Edd}}\sim 0.93$, exhibits continuous growth in a single super-Eddington peak without intermittent drops.  
Similarly, the absence of fluctuations in the \texttt{Therm} model, akin to the \texttt{NoAGN} model, indicates the relatively limited effect of thermal heating in our models compared to the more impactful kinetic winds and radiative components.

For SMBHs with masses above $5\times10^7\,M_{\odot}$, the accretion rate in the \texttt{Edd\_lim} model is higher than the $\lambda_{\mathrm{Edd}}$ inferred from observations.
This indicates that the steady growth in the Eddington-capped model leads to a BH mass-accretion rate relation at odds with high redshifts observations. 
For these massive BHs, the results of our AGN feedback-regulated models are constrained by the mass of the host galaxy in this work. However, given the high BH mass-accretion rates reached by these feedback-regulated models, applying our complete AGN feedback to a more massive galaxy could likely yield BH masses exceeding $5\times10^7\,M_{\odot}$.
%

\subsection{The effect of black hole offsets with respect to the host galaxy on its evolution}\label{sec:pos}


In Sec.~\ref{sec:smbh}, we discussed how BH mass growth is shaped by the properties of gas available for accretion {--} primarily limited by both AGN and SNe feedback.
However, the dynamics of SMBHs can also influence their growth and accretion rates \citep{2018MNRAS.480.3762S,2019MNRAS.486..101P}.
In this section, we explore how the BH moves within its hosting galaxy to better understand changes in its accretion rate over time. 
For studying the BH offset, a key advantage of the simulations in this work is that the BH trajectory is determined directly by the gravitational field solution, without any artificial constraints pinning the BH to the region of maximum density or to the center of the halo.   

Figure~\ref{fig:sinkpos} shows the distance of the sink particle from the center of its host galaxy, $\Vert{r_{\mathrm{sink}}}\Vert$,  over time.
Initially, the BH seed forms at the exact position where its progenitor star dies. Since this is the first star to form in the simulation box, the BH initial position closely coincides with the peak density of the galaxy (Sanati et al \textit{in prep.}).  
Over time, both the BH position and the 
center of mass of the galaxy shift, as revealed by their varying relative distances in the figure.
%
%
%
%

%
%

We find different feedback models to have an important effect on the relative BH displacement.
In the \texttt{NoAGN}, \texttt{Edd\_lim} and \texttt{Therm} models, featuring either no AGN feedback or an inefficient only-thermal feedback,
$\Vert{r_{\mathrm{sink}}}\Vert$ can reach values as small as $\sim1\mathrm{pc}$.
In the \texttt{NoAGN} and \texttt{Therm} models, the BH spends $\lesssim 15\,\mathrm{Myr}$ inside the central region. During this time, it  grows rapidly, reaching a maximum accretion rate of $\simeq14\,{\dot {M}}_{\mathrm{Edd}}$. 
This initial phase of fast accretion leads to significant BH mass growth, which increases the dynamical friction it experiences.  As a result, the BH remains anchored near the center, enabling continued growth.
However, as discussed in Sec.~\ref{sec:smbh}, this rapid accretion also depletes the surrounding gas, weakening the central gravitational potential.
As a result, after this brief phase of high accretion near the galaxy center, the BH struggles to stay within $\simeq10\,\mathrm{pc}$ of the central region. 
%
%
There is a similar interplay between BH mass growth, and its trajectory when comparing the \texttt{ThermKin}, \texttt{ThermKinRad}, \texttt{ThermHKin}, and \texttt{ThermHKinRad} models to the \texttt{Edd\_lim} model. For the BH to grow efficiently in the feedback-regulated models, its trajectory must intersect dense gas clouds, which are primarily concentrated in the galactic center. However, in the feedback-regulated runs, the BH typically maintains an average distance of a few hundred $\mathrm{pc}$ from the center, while in the \texttt{Edd\_lim} model it stays closer, within $\lesssim100\,\mathrm{pc}$, for most of its evolution. This difference arises because the lighter BHs in the feedback-regulated models experience weaker dynamical friction and are more easily perturbed, especially during merger events, compared to the more massive, steadily growing BH in the \texttt{Edd\_lim} model.

%


\begin{figure}
    \centering
    \includegraphics[width=0.48\textwidth]{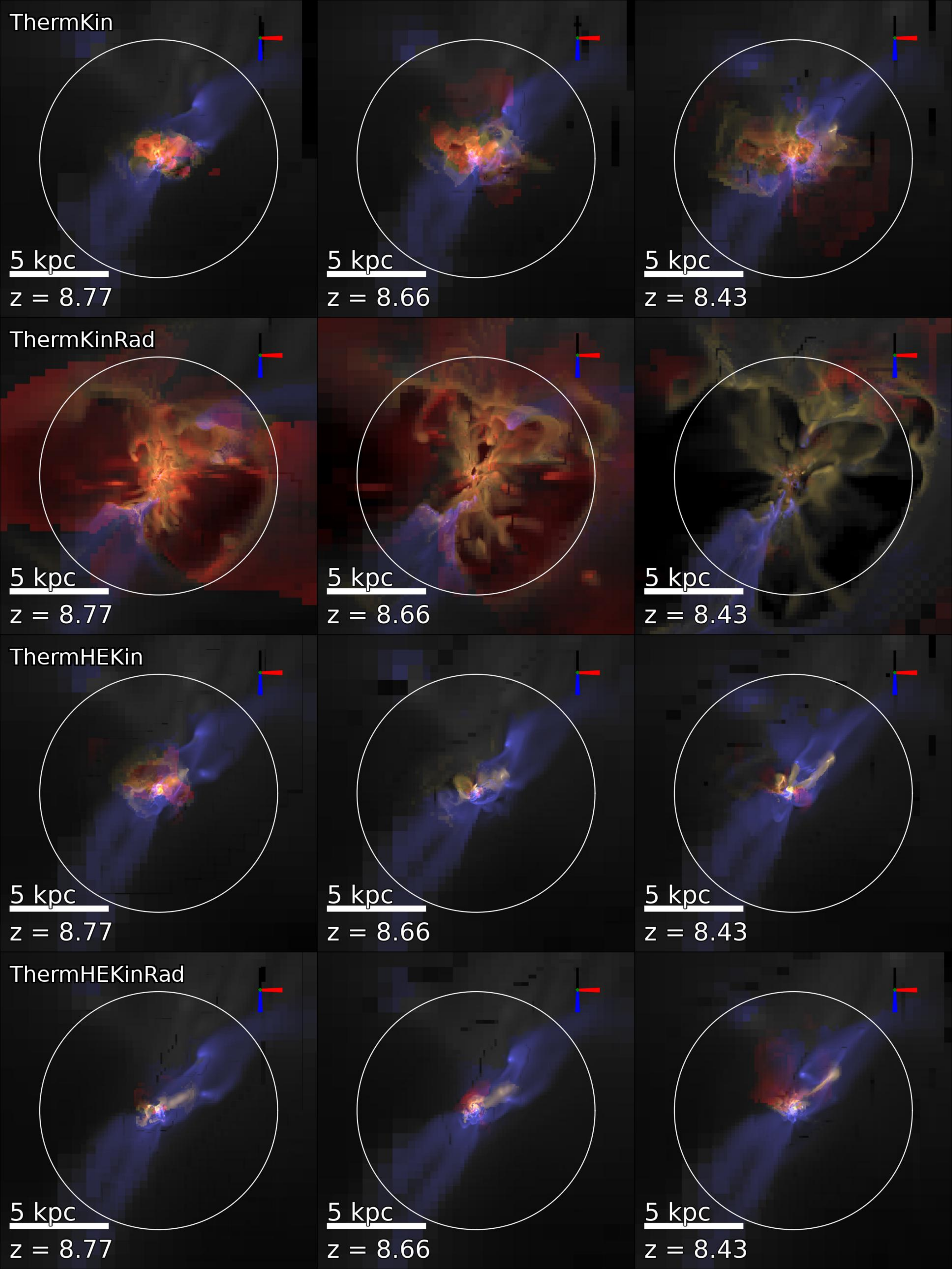}

    \caption{Density-weighted projection of gas density in gray, separated into inflowing gas in blue with radial velocity $v_{r}<-30\,\mathrm{km\,s^{-1}}$, and high speed outflowing gas in red with $v_{r}>300\,\mathrm{km\,s^{-1}}$, and low speed outflowing gas in green with $v_{r}>30\,\mathrm{km\,s^{-1}}$. In the \texttt{ThemKinRad} model with strong accretion rate, AGN-driven radiation pressure generate outflows extending to $\sim50\,\mathrm{kpc}$ at redshift $z\sim8.5$ with maximum radial velocity of exceeding $2500\,\mathrm{km\,s^{-1}}$. The white circle shows the galactic region with radius of $6\,\mathrm{kpc}$ at redshift $z \sim 8$ which corresponds to $0.15$ times the Virial radius of the galaxy.}

    \label{fig:outflows}
\end{figure}
\begin{figure}
    \centering

    \includegraphics[width=0.48\textwidth]{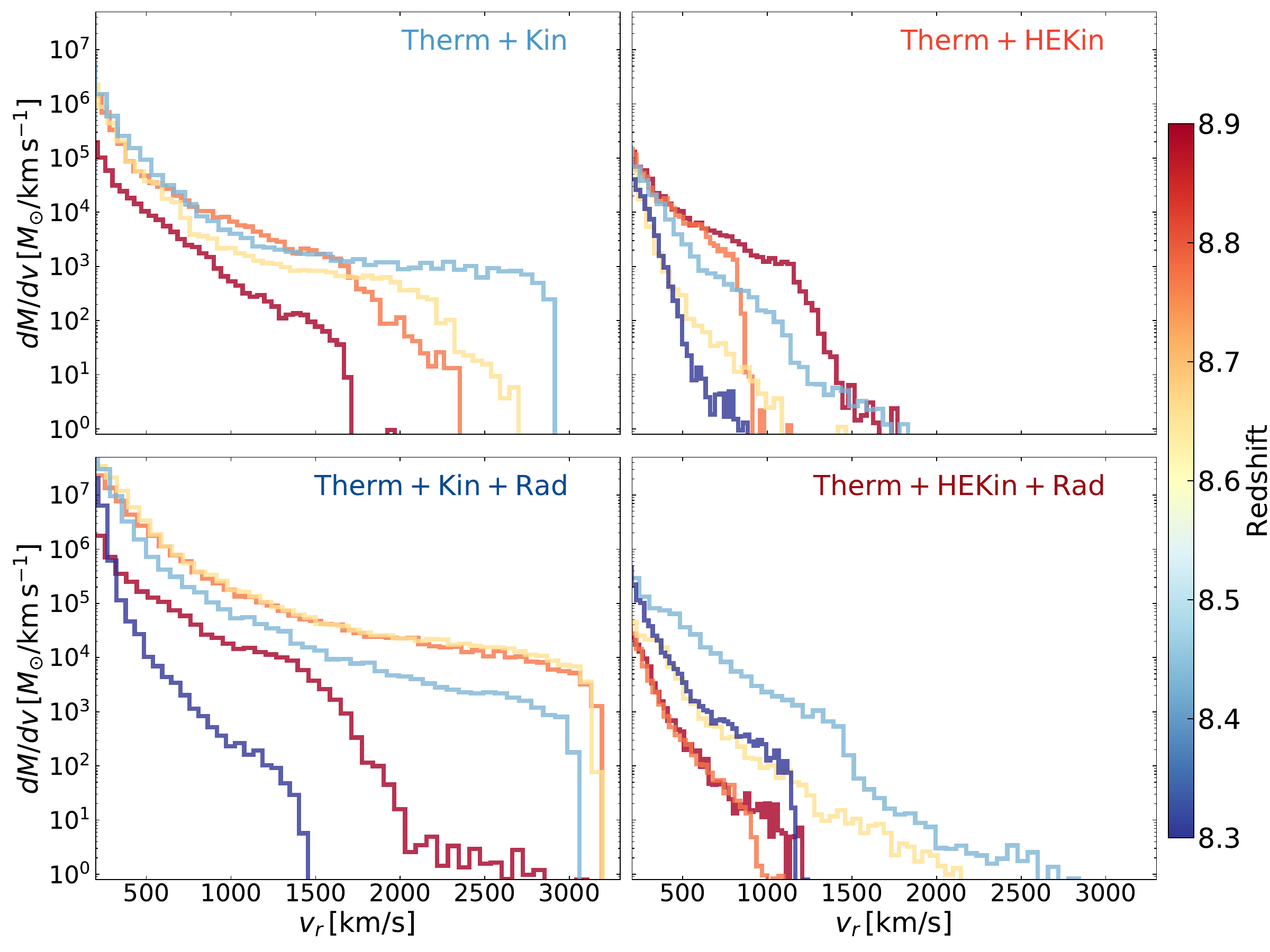}
    \caption{Radial velocity distribution of outflowing gas within a physical radius of $6\,\mathrm{kpc}$ (equivalent to the white circle in Figure.~\ref{fig:outflows}). 
    The figure focuses on $71\,\mathrm{Myr}$ of outflow evolution between redshift $z\sim8$ and $9$, following the merger event at redshift $z\sim9$ and subsequent cold gas inflow. Each panel represents a different AGN-feedback regulated model.
    Strong outflows with maximum radial velocities exceeding $2000\,\mathrm{km\,s^{-1}}$ in model \texttt{ThermKinRad} arise from the combined effects of high BH accretion rates and AGN-driven radiation pressure, 
    which we attribute to radiation unbinding the dense gas surrounding the BH.}
    \label{fig:vr}
\end{figure}

\begin{figure*}
    \centering
    \includegraphics[width=\textwidth]{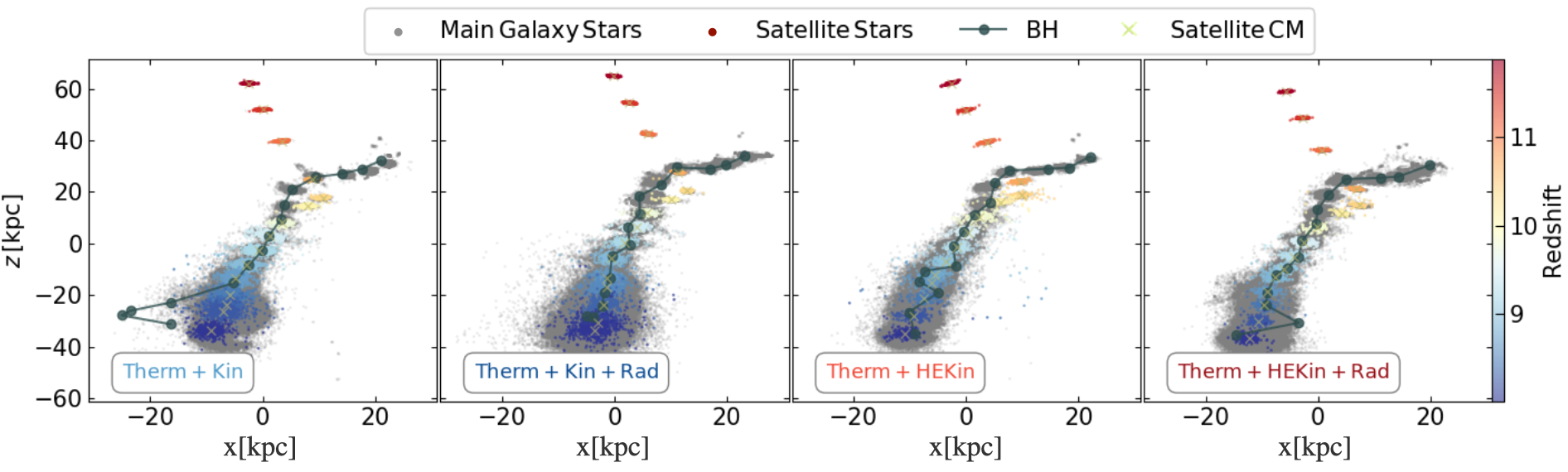}


    \caption{Merger event between redshifts $z\sim 11$ and $8$. The panels from left to right show the \texttt{ThermKin}, \texttt{ThermKinRad}, \texttt{ThermHKin}, and \texttt{ThermHKinRad} models, respectively. The redshift range is selected to showcase the stellar distribution in the main halo of the galaxy hosting the BH, as well as position of the BH, before and after the infall of its satellite. The stars in the main galaxy are shown in gray, while those in the satellite are color-coded with the corresponding redshift. The green circle and the yellow cross show the position of the BH and the center of mass of the satellite galaxy, respectively}
    \label{fig:stellar_mass}
\end{figure*}

\subsection{AGN-driven galactic scale outflows}\label{sec:outflows}

Figure~\ref{fig:outflows} showcases the qualitative properties of the outflows in the AGN-regulated models, at increasing time from left to right columns. 
The density-weighted projection of the total gas density is shown in gray. Inflowing gas with a radial velocity offset with respect to the galaxy of $v_{r}<-30\,\mathrm{km\,s^{-1}}$ is shown in blue. Outflowing gas is separated into two groups: slow winds with $v_{r}>30\,\mathrm{km\,s^{-1}}$ shown in orange, and fast winds with $v_{r}>300\,\mathrm{km\,s^{-1}}$ shown in red. The white circle shows the galactic region with size $\sim6\,\mathrm{kpc}$, equal to $0.15$ of the virial radius of the host dark matter halo. 
From top to bottom, panels in the first and third rows show weak and strong radio feedback, \texttt{ThermKin} and \texttt{ThermHKin}, respectively. The second and fourth rows show the \texttt{ThermKinRad} and \texttt{ThermHKinRad} models, which also include AGN 
radiative feedback
in addition to radio modes.

At redshift $z\sim9$, the BH undergoes a period of intense accretion, driven by a merger event.
This elevated accretion rate generates strong AGN feedback.
%
%
In the \texttt{ThermHEKinRad} model (forth row), the BH accretion rate is the lowest among the four models, resulting in weak AGN feedback and the smallest fraction of outflowing gas (in red).
The \texttt{ThermKin} and \texttt{ThermHEkin} models (first and third rows), BH has a higher accretion rate and generates stronger feedback (see Sec.~\ref{sec:smbh}). However, the left-most panels at $z=8.77$ show that the accreting BH remains enshrouded by dense gas layers (in red), causing the feedback to become confined within the galactic nucleus.
Since thermal energy is included in all of these models, this confinement shows that thermal pressure alone is insufficient to lift against the potential the gas layers enshrouding the AGN. This is a result of the high binding energy of the central region.
In contrast, in the \texttt{ThermKinRad} model (second row), after an event of high accretion rates, AGN-driven radiation pressure generates outflows extending to a comoving radius of $\sim50\,\mathrm{ckpc}$ at redshift $z\sim8.5$.
This is due to two key factors: 
1) A comparison between the high accretion rates in the \texttt{ThermKin} and \texttt{ThermKinRad} models at redshift $z\sim9$ (see Fig.\ref{fig:BHStGs_acc}), following the merger event, shows that a high BH accretion rate alone is not sufficient to sweep up the gas through feedback.
Although in both models BH undergoes a super-Eddington accretion at this redshift, the momentum-driven component of the feedback is too weak to unbind the tightly held gas and initiate a significant outflow.
This highlights the critical role of AGN radiation feedback in driving galactic-scale outflows.
%
2) A comparison between the \texttt{ThermKinRad} and \texttt{ThermHKinRad} models shows that radiation pressure is particularly effective at generating galactic outflows when the BH accretion rate is comparable to the Eddington limit.
While both models include radiative feedback, the \texttt{ThermKinRad} model shows significantly stronger outflows due to its more efficient BH growth. The accumulation BH mass in this model is approximately $50$ times higher than in the \texttt{ThermHKinRad} model (see Sec.~\ref{sec:smbh}), resulting in stronger AGN feedback.

In summary, AGN radiative feedback enables the BH to have a greater impact on the hist galaxy, by allowing for more efficient and rapid BH growth, primarily through the suppression of star formation, (as discussed in Sec.~\ref{sec:smbh}), and by driving strong outflows. In the latter process, radiation heats and rarefies the surrounding gas, as the kinetic winds can scape through the low density channels, creating low-density channels through which kinetic winds can more easily escape.
We note that in massive galaxies, hosting $> 10^9\,M_\odot$ SMBHs, radiation has a limited impact and a multi-scattered radiation pressure in the form of infrared (IR) feedback is required to obtain fast outflows \citep{costa_quenching_2018}. 
Although IR feedback resulting from radiation pressure on dust is not included in this work, its impact is more significant 
in more massive $z>6$ galaxies with high metal enrichment. The small mass of the galaxy simulated in this work, combined with its low IRoptical depth due to its low metallicity, minimizes the relevance of IR feedback in this context.

Disentangling the physical mechanisms that drive galactic outflows is observationally challenging due to the complex interplay between different feedback processes.
%
Supernova feedback may contribute to creating low-density chimneys within the central dense gas, making it easier for outflows to launch. It can have a complementary effect to AGN feedback in expelling gas. 
To verify that the galactic-scale outflows, particularly in the \texttt{ThermKinRad} model, are primarily AGN-driven, with stellar feedback playing a secondary role, we compare the gas velocity in this model to that in others. This allows to assess the relative contributions of AGN and SNe feedback in driving the outflows.
Figure~\ref{fig:vr} shows the probability distribution function (PDF) of radial velocity across different models. 
The radial velocity is computed for  gas located within $\sim6\,\mathrm{kpc}$ physical radius of the simulated galaxy.
This radius is equivalent to the white circle in Fig.~\ref{fig:outflows}. 
We focus on $\sim71\,\mathrm{Myr}$ of outflow evolution, between redshift $z\sim9$ and $8$. 
This marks the period right after the BH host galaxy has merged with a satellite companion 
(see also Sec.~\ref{sec:merger}). 
The different line colors represent successive time steps, with red corresponding to higher redshift ($z\simeq9$) and blue to later times ($z\simeq8$). At early times (red curves), there is little high-velocity outflowing gas in any of the models. As time progresses, particularly around $z\simeq8.5$, shown in light blue, yellow, and orange, high-velocity outflows begin to develop. This is especially prominent in the \texttt{ThermKinRad} model, which shows a significant broadening of the PDF toward higher velocities. Eventually, as the gas expands and escapes, the overall radial velocity distribution shifts and narrows again (dark blue).
%
%
While the BH accretion rate increases in all models after the merger event, $\dot{M}_{\mathrm{BH}}$ in \texttt{ThermKinRad} model reaches $5-15$ times higher compared to the rest.
%
%
After the merger event, the star formation rate increases similarly across all models, with a peak at $z\simeq9$. However, the high radial gas velocities, exceeding $2500\,\mathrm{km\,s^{-1}}$, correlates more closely with a significant boost in BH accretion rate in the \texttt{ThermKinRad} model, rather than with increased star formation activity present across all models, following the merger event, whose redshift is marked by the gray dashed lines.  
Although all models exhibit outflows with velocities $v_{\mathrm{r}}>500\,\mathrm{km\,s^{-1}}$, the fraction of gas reaching such speeds is $\sim400$ times greater in the \texttt{ThermKinRad} model. This aligns with high-redshift observations of outflows at similar velocities \citep{carniani_jades_2024,2025arXiv250117145S}, suggesting that a significant mass of high-velocity gas is required for detection. 
In summary, the high velocities of the outflows, specially considering their occurrence following a significant BH growth, supports the argument that these outflows are primarily AGN-driven rather than resulting from SNe feedback. 
Such energetic outflows, when breaking out of the galaxy, carry magnetic energy along their path \citep[e.g.,][]{2017MNRAS.464.4448W,garcia_magnetization_2021,blunier_constraint_2024}. The extent of these magnetised outflows, generated by AGN feedback, can be compared to the remnants of primordial magnetic fields from the early universe \citep[e.g.,][]{wasserman1978,1996ApJ...468...28K,gopal_large_2003,subramanian_origin_2016,sanati_constraining_2020,garg_are_2025}. 
This comparison could reveal the role of low-mass galaxies in magnetising the IGM, particularly inside and at the edges of cosmic voids where dwarf-size galaxies are more abundant and their outflows may play a more significant role than those of more massive galaxies. 




\begin{figure}
    \centering

    \includegraphics[width=0.48\textwidth]{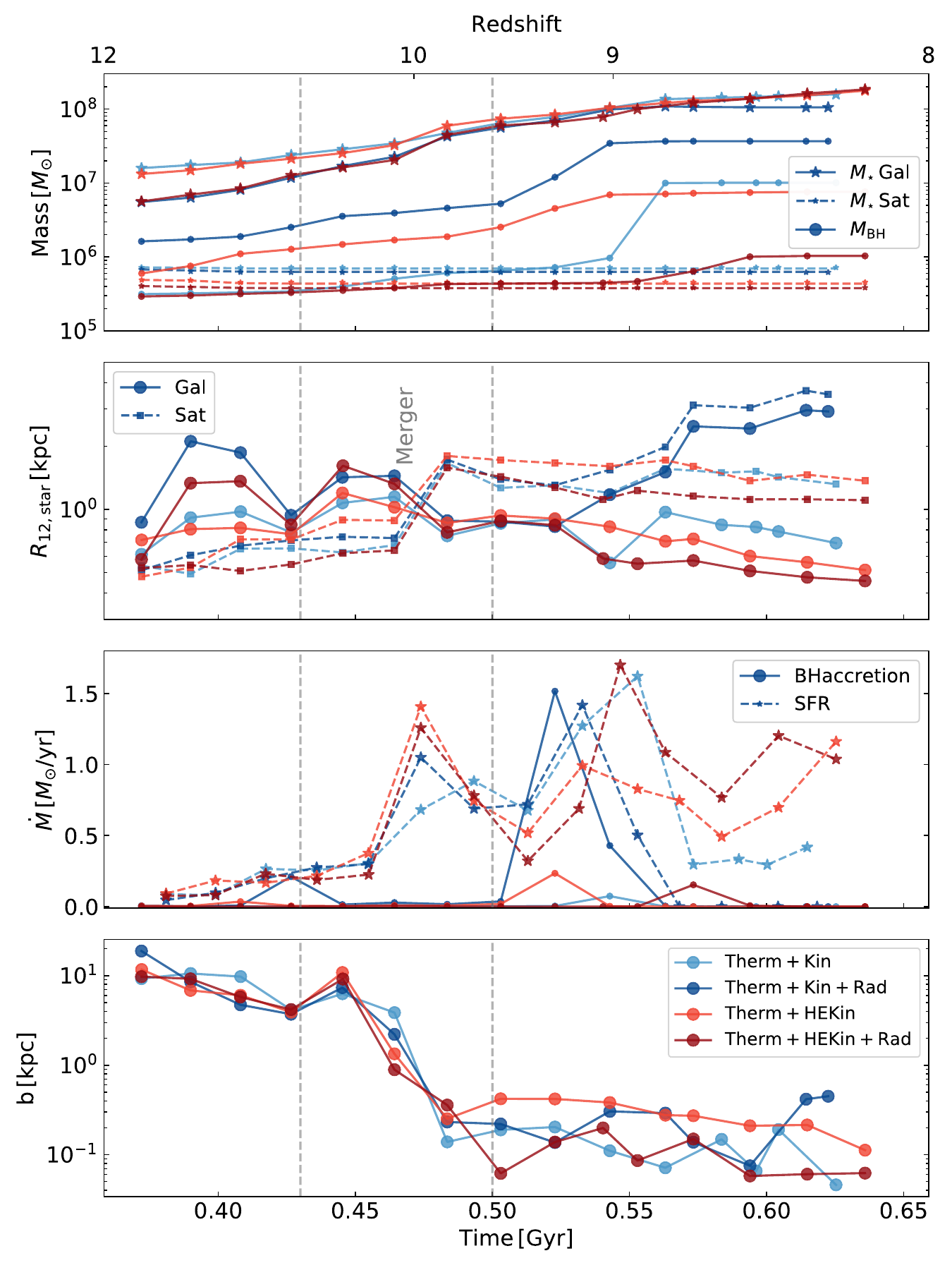}
    \caption{Galaxy and BH properties, before and after a merger event between redshift $z\sim 11$ and $8$, in feedback-regulated models. 
    First panel) The stellar mass of the main galaxy hosting the BH, and that of its satellite are shown by stars connected by solid and dashed lines, respectively. Circles connected by solid lines indicate a sharp increase in BH mass across all models. 
    Second panel) The half mass radius, $R_{12}$, of the main galaxy and its satellite are shown by solid and dashed lines, respectively. The merger event disturbs the stellar distribution in the main galaxy, but the AGN outflows in the \texttt{ThermKinRad} model cause a further increase in $R_{12}$.
    Third panel) The star formation rate (SFR) and BH accretion rate. During the merger event, gas compression boosts the SFR. A second SFR peak results from the inflow of cold gas from the satellite. BH accretion rates increase in all models post-merger, with a prominent peak in the \texttt{ThermKinRad} model.
    Fourth panel) The merger impact 
    shows a similar merger process across all models. 
    }
    \label{fig:merger}
\end{figure}

\subsection{Impact of merger on BH growth}\label{sec:merger}

As discussed in Sec.~\ref{sec:smbh}, one possible trigger for efficient accretion onto a central galaxy BH is a merger event. Such event enables clumps of cold, dense gas to refill the innermost regions of the BH host galaxy, 
driving significant BH growth following the merger.  
The merger is illustrated in Fig.~\ref{fig:stellar_mass}. 
At redshift $z<12$, a satellite galaxy with a mass of  $\sim5\times10^5\,M_{\odot}$ approaches the BH host galaxy, which has a stellar mass of $M_{\star}\sim10^7\,M_{\odot}$. The two galaxies are initially at a distance of $\gtrsim10\,\mathrm{kpc}$ at redshift $z\simeq12$. 
The infall process concludes before redshift $z=8$. Stellar particles in the host galaxy are shown in gray.
The satellite galaxy is color-coded according to redshift, with its center of stellar mass marked by a yellow cross.

Before the merger, the BH host galaxy has a size of $\sim\,1\,\mathrm{kpc}$ at redshift $z\simeq12$.
After the merger, although the galaxy remains intact, its stellar distribution is disturbed by the satellite infall, which leads to an increase in its size across all models.  
The origin of this size growth, occurring over $\sim 250\,\mathrm{Myr}$ during the merger is driven by the expansion of the stellar component and is likely linked to the rapid modification of the gravitational potential \citep{2012MNRAS.421.3464P, 2013MNRAS.432.1947M}. 
The dark green circle shows the position of the sink particle, traced by solid lines.
While the infall of the satellite supplies the BH host galaxy with approximately $10^6\,M_{\odot}$ of gas at redshift $z\simeq12$, this supply is secondary to the mass already present in the system. Instead, the primary effect of the merger on the BH growth is inducing the compression of mass and gravitational potential in the central region, leading to enhanced BH accretion.
Second, it disturbs the BH dynamics, and shifts its position relative to the galaxy center. 


%

%

Figure~\ref{fig:merger} more quantitatively shows the properties of both BH and galaxy during the merger. 
%
The top panel shows the stellar mass of the main galaxy hosting the BH, and that of its satellite, with star symbols connected by solid and dashed lines, respectively.
Before the merger, the stellar mass in the BH host galaxy varies from $M_{\star}=\,5\times10^6$ in the \texttt{ThermHKinRad} to $2\times10^7\,M_{\odot}$ in the \texttt{ThermKin}. 
Interestingly, the stellar masses in the two models with AGN radiation are similar, and lower than those in the two models without this component, which also overlap. This shows the impact of AGN radiation feedback in reducing the star formation rate. 
%
The merger happens between this galaxy and a smaller companion with stellar mass of $\sim4-8\times10^5$,
which is slightly lower in models with stronger AGN feedback. 
The stellar masses in the companion galaxy follow this order: \texttt{ThermKin}, \texttt{ThermKinRad}, \texttt{ThermHKin}, and \texttt{ThermHKinRad}. 
While the role of stochasticity in galaxy evolution simulations cannot be overlooked, this difference in the stellar mass may be due to the long-range effects of AGN feedback on star formation in neighboring galaxies, extending up to several $\mathrm{kpc}$ (The initial distance between the two galaxy is $\gtrsim10\,\mathrm{kpc}$ at redshift $z\simeq12$).
The evolution of BH mass during the merger event is depicted with circles connected by solid lines, showing a sharp increase in BH growth across all models, immediately following the infall of the satellite galaxy. This highlights the dynamic interplay between galactic interactions and BH evolution, particularly in the more chaotic environment of high redshifts.

%

The merger also disrupts the stellar distribution in both the host and satellite galaxies. 
The impact on the half mass radius, $R_{12}$, is shown in the second panel of Fig.~\ref{fig:merger} for both galaxies. 
Prior to the merger, the size of the satellite galaxy is relatively similar across different feedback models.  However, $R_{12}$  in the main galaxy shows slight variations between the models.  
These differences appear to correlate with the strength of 
the AGN feedback, with the galaxy being more extended when it is hosting a more massive BH that generates stronger feedback 
(see also Sec.~\ref{sec:outflows}). 
Following the merger, there is an in situ burst of star formation. With all newly formed stars associated to the main galaxy. These stars are primarily formed in the central region of the system, effectively decreasing its half-mass radius and leading to a more compact galaxy.
An exception is the \texttt{ThermKinRad} model, where strong AGN radiative feedback dominates and disperses the stars instead. 
In all feedback models, the merger disturbs and gravitationally heats the stellar distribution, causing the size of the satellite galaxy to increase.

%


In the third panel we examine the impact of the merger on both the BH accretion rate and the star formation rate (SFR) in the host galaxy, represented by solid and dashed lines, respectively.
The SFR begins to rise sharply at redshifts $z\lesssim11$ across all models. 
We attribute this to gravitational interactions between the merging galaxies, which compress the gas and enhance the 
star formation efficiency, as well as the overall growth in halo mass. 
%
%
A similar trend is observed for the BH accretion rate, which also experiences a significant boost following the merger. 
Among the feedback regulated models here, the \texttt{ThermKinRad} model shows the highest accretion rate peak.
$\dot{M}_{\mathrm{BH}}$ in this model reaches $1.5\,M_{\odot}\,\mathrm{yr}^{-1}$, compared to $0.1-0.3\,M_{\odot}\,\mathrm{yr}^{-1}$ in the other models.
This is consistent with its hosting of the most massive BH before the merger, as a larger BH mass naturally supports a higher accretion rate. 

In the bottom panel, we analyze the characteristics of the infall by examining the impact parameter of the merging galaxies. The impact parameter decreases over time as the satellite galaxy approaches the central galaxy. A comparison across feedback-regulated models reveals that the BH host galaxy undergoes a similar merging process, with comparable gravitational interactions in all cases. Thus the observed differences in BH growth, accretion rate and sftar formation, are likely attributed to the distinct effects of AGN feedback on the host galaxy in each model.

\begin{figure}
    \centering

    \includegraphics[width=0.48\textwidth]{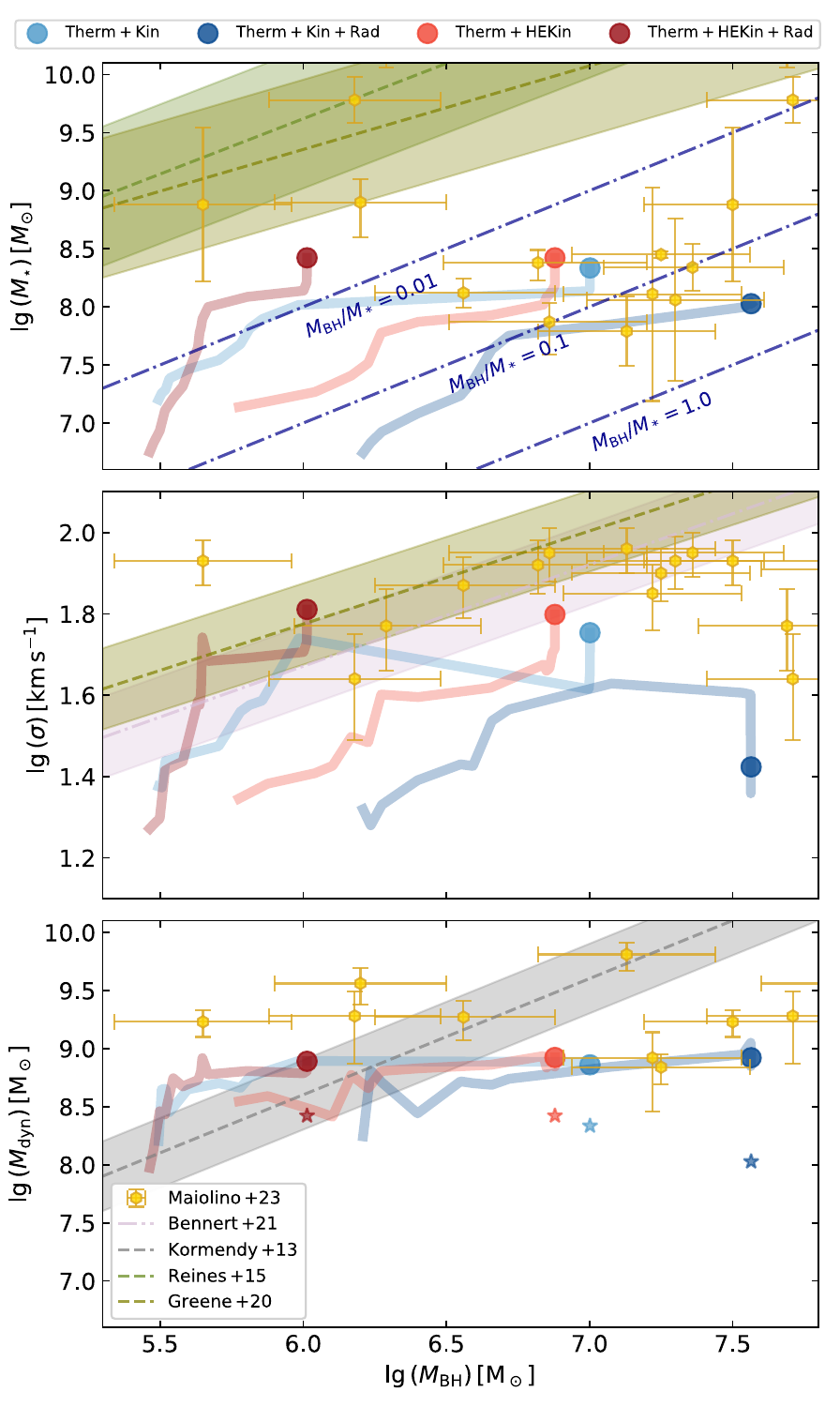}
    \caption{Host galaxy properties as a function of BH mass in AGN feedback-regulated models.
    The panels show the time evolution of stellar mass (top), stellar velocity dispersion (middle), and stellar dynamical mass (bottom) as a function of BH mass, between redshift $z\sim12$ and $8$. The results are compared with JADES observations \citep{maiolino_jades_2023}, in yellow hexagons, and local scaling relations of \citet{2021ApJ...921...36B}, \citet{kormendy_coevolution_2013}, \citet{2015ApJ...813...82R}, and \citet{2020ARA&A..58..257G} in dashed pink, gray, light green and dark green lines, respectively. Shaded regions show the dispersion and uncertainty of the fits. 
    In simulated galaxies, the $M_{\mathrm{BH}}/M_{\star}$ ratio remains between $0.01$ and $1$, as shown by the dark blue dotted lines. The exception is the \texttt{ThermHKinRT} model with the lowest BH mass growth and $M_{\mathrm{BH}}/M_{\star}$ ratio below $0.01$.
    Overall, the scaling relations show a great agreement with high-redshift observations.
    }
    \label{fig:Mdyn}
\end{figure}

\subsection{BH-host galaxy scaling relations}
Figure~\ref{fig:Mdyn} shows the co-evolution of the host galaxy properties with the mass of their hosted BH. 
%
The top panel shows the evolution of stellar mass, represented by solid lines, with circles indicating the final stellar at redshift $z=8$. 
Straight dashed-dotted lines in dark blue correspond to constant values of $M_{\mathrm{BH}}/M_{\star}=0.01$, $0.1$, and $1$.
Local scaling relations obtained by \citet{2015ApJ...813...82R} and more recently by \citet{2020ARA&A..58..257G} are shown in dark and light green, respectively.
Dashed lines show their best fitting relation, with the shaded area indicating the scatter and slope uncertainty in each sample. 
The yellow hexagons show the JADES observations of quasars at redshift $4<z<11$ \citep{maiolino_jades_2023}.
The JADES sample exhibits a significant deviation from local observations in the BH mass–stellar mass scaling relation, particularly for low stellar mass galaxies with $M_{\star}<10^9\,M_{\odot}$. 
The stellar mass vs black hole mass relation of our simulated galaxy is comparable to the outliers in the JADES sample, and we observe a similar deviation from the low-redshift scaling relation at high redshift. 
Consistent with high-redshift observations, all our feedback-regulated models show that BHs between redshifts $z \sim 12$ and $8$ are significantly over-massive relative to the stellar mass of their host galaxies, compared to local scaling relations. In the \texttt{ThermKinRad} model, we even find the black hole mass approaches 
a significant fraction of its host galaxy stellar mass.

The middle panel shows the mass-weighted 3D stellar velocity dispersion $\sigma$. 
As in the top panel, the green shaded region represents the $1$ sigma dispersion of the Local scaling relation provided by \citet{2020ARA&A..58..257G}. 
For calculating $\sigma$, we use the stellar component within $2\,R_{1/2}$, where $R_{1/2}$ is the half stellar mass radius.
%
%
The evolution of the \texttt{ThermHKinRad} model (red line), with the strongest AGN feedback and least efficient BH growth, 
aligns with the higher velocity dispersion values inferred from observations in the Local Group. 
However, we note that at such low stellar masses, the local scaling relation is not well constrained and primarily relies on extrapolation from the higher mass regime.
%
The three other models with more efficient BH growth deviate from the Local $M_{\mathrm{BH}}-\sigma$ relation and instead are more closely consistent with high-redshift JWST observations.


The bottom panel shows the dynamical mass of the galaxy, $M_{\mathrm{dyn}}$, as a function of its residing BH mass. 
The gray dashed line shows the best fit to the BH mass-bulge mass data from \citet{kormendy_coevolution_2013}, and the shaded region shows the $1$ sigma dispersion of the relation. 
Same as velocity dispersion, the stellar $M_{\mathrm{dyn}}$ in our simulations is computed within twice the half-mass radius, and linked to the velocity dispersion through the Virial theorem, 
\begin{eqnarray}
    M_{\mathrm{dyn}} = \frac{2\,R_{1/2}\times\sigma(2\,R_{1/2})}{G},
\end{eqnarray}
To facilitate comparison, the final stellar mass in each model at redshift $z\simeq8$ is shown with stellar symbols. As expected, the dynamical masses in all models are higher than the stellar masses, and the deviation from the local fit is not as strong as the $M_{\mathrm{BH}}-M_{\star}$ scaling relation. 

In summary, the SMBH-galaxy scaling relations obtained for the modeled galaxies indicate that black hole seeds grow efficiently, reaching masses comparable to or exceeding $10^7\,M_{\odot}$ by $z=8$, while maintaining a black hole to stellar mass ratio within  $0.01<M_{\mathrm{BH}}/M_{\star}<1$.
The SMBHs mass, along with the stellar and dynamical masses of the host galaxy, is consistent with high-redshift observations of moderate-luminosity quasars in low-mass galaxies. However, similar to these observations, the black holes in our models appear overmassive relative to the stellar mass of galaxies they reside in as predicted by the SMBH-galaxy scaling relation in the local Universe.

%


\section{Conclusions and discussion}\label{sec:sum}

In this work, we explore how various AGN physical processes and configurations shape the co-evolution of a model galaxy and its central SMBH at high redshifts.
We conduct cosmological radiative-magneto-hydrodynamical simulations of a zoom-in galaxy with stellar mass of $\simeq5\times10^8\,M_{\odot}$ down to redshift $z=8$. All simulations are generated using our modified version of the \textsc{Ramses} code.
%
%
%
In addition to two models without AGN feedback and one where black hole accretion is limited by the Eddington rate, we systematically build up the AGN feedback by adding different physical components and processes. These include thermal energy injection, kinetic winds (representing both weak and strong radio-mode feedback), and AGN radiation. This step-by-step approach allows us to deconvolve and study how each feedback mechanism contributes to regulating black hole accretion and growth.
In these simulations, black hole seeds originate from Population III.1 star progenitors. 
We self-consistently model the formation and irradiation of the progenitor star, 
explicitly capturing preheating effects often neglected in simulations seeding SMBHs.  
We believe that our results are therefore also relevant to other SMBH formation pathways that involve preheating of the gas in the host halo by internal or external sources.
%
%
%
%
%
Our main findings are as follows:

\begin{itemize}

    \item  
    We find SMBHs to rapidly grow during the dawn of the first galaxies, through AGN self-regulated accretion. Our $10^5\,M_\odot$ initial mass seeds can grow and even exceed $10^7\,M_\odot$ by redshift $z=8$, with black hole to stellar mass ratio of  $0.01<M_{\mathrm{BH}}/M_{\star}<1$. With typical host galaxy stellar masses of $\gtrsim10^8\,M_{\odot}$, our simulated SMBHs resemble high-redshift observations of moderate-luminosity quasars fueled by overmassive SMBHs in low-mass galaxies.  
    %
    %
    %
    
    \item 
    %
    We find super-Eddington accretion episodes, reaching up to approximately hundred times the Eddington limit.
    %
    These episodes 
    are driven either by continuous gas inflows or triggered by merger events between the BH host halo and the neighboring subhalos, mainly at redshifts between $z=12$ and $8$.
    These episodes are followed by strong AGN feedback that thwarts black hole growth, driving a feedback regulating cycle that prevents sustained super-Eddington accretion.

    \item
    We find AGN radiative feedback to play a crucial role in launching high-velocity galactic outflows. 
    It creates low density channels that allow feedback to escape when high-density gas accumulates around the BH during super-Eddington accretion events. This drives strong outflows extending to $\sim50\,\mathrm{kpc}$ at redshift $z\sim8.5$, featuring maximum radial velocities exceeding $2500\,\mathrm{km\,s^{-1}}$.


    \item 
    The SMBH are allowed to move freely within our simulations, with them typically remaining at a few hundred $\mathrm{pc}$ from the galaxy center. We find that the BH mass growth and its proximity to the galaxy center correlates over its evolution. 
    %
      
    \end{itemize}

Overall, our work shows the important role of AGN feedback in regulating the growth of SMBHs, and the importance of accounting for additional physical processes such as their radiative feedback. Variations in these physics shape the co-evolution of SMBHs with their host galaxy as well as the star formation rate and gas dynamics within galaxies.
Moreover, AGN feedback are closely linked to the launching of high-velocity outflows that can reach large distances, injecting metals and energy into the IGM (Sanati et al in prep.), and thereby influencing a broader region than the galaxy nucleus.
In future work, we will investigate the effects of accounting for cosmic rays and magnetism. These are particularly poorly understood in the context of high-resolution galaxy formation. They may influence galactic-scale outflows, cooling rates, and the propagation of energy through the ISM and IGM.
We anticipate them to have a non-negligible impact on gas dynamics in the immediate vicinity of the BH as well as on star formation in the galactic scale. Investigating these additional components will be crucial to gaining a more comprehensive understanding of the interplay between AGN feedback, BH growth, and galaxy evolution in a cosmological context.

\section*{Acknowledgements}
M.S. acknowledges the support from the Swiss National Science Foundation under Grant No. P500PT\_214488 and the Chalmers Initiative on Cosmic Origins (CICO) postdoctoral fellowship.
J.C.T. acknowledges support from ERC Advanced Grant 788829 (MSTAR) and the
CCA Sabbatical Visiting Researcher program.
This work used the DiRAC@Durham facility managed by the Institute for Computational Cosmology on behalf of the STFC DiRAC HPC Facility (www.dirac.ac.uk). The equipment was funded by BEIS capital funding via STFC capital grants ST/P002293/1, ST/R002371/1 and ST/S002502/1, Durham University and STFC operations grant ST/R000832/1. DiRAC is part of the National e-Infrastructure. This work was performed using resources provided by the Cambridge Service for Data Driven Discovery (CSD3) operated by the University of Cambridge Research Computing Service (www.csd3.cam.ac.uk), provided by Dell EMC and Intel using Tier-2 funding from the Engineering and Physical Sciences Research Council (capital grant EP/P020259/1), and DiRAC funding from the Science and Technology Facilities Council (www.dirac.ac.uk).
S.M.A. is supported by a Kavli Institute for Particle Astrophysics and Cosmology (KIPAC) Fellowship, and by the NASA/DLR Stratospheric Observatory for Infrared Astronomy (SOFIA) under the 08\_0012 Program. SOFIA is jointly operated by the Universities Space Research Association, Inc. (USRA), under NASA contract NNA17BF53C, and the Deutsches SOFIA Institut (DSI) under DLR contract 50OK0901 to the University of Stuttgart.

\section*{DATA AVAILABILITY}
The data employed in this manuscript is to be shared upon reasonable request contacting the corresponding author.

\bibliographystyle{mnras}
\bibliography{bibliography, references}

\appendix
\section{SMBH accretion rate}\label{sec:app}

\begin{figure}
    \centering
    \includegraphics[width=0.48\textwidth]{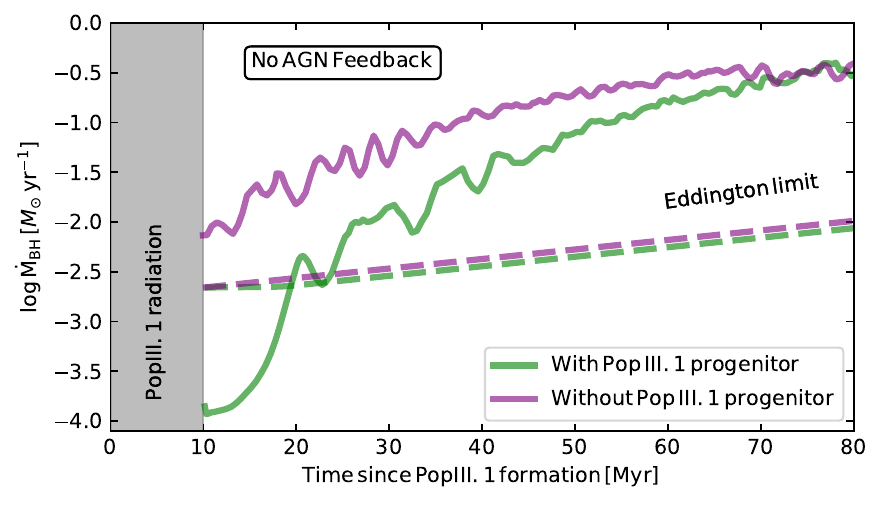}

    \caption{Evolution of the black hole accretion rate following the formation of a SMBH from a Pop III.1 progenitor in green and without a Pop III.1 progenitor in purple. The solid and dashed lines show the Bondi-Hoyle and Eddington accretion rates, respectively. 
    The Pop III.1 star emits a strong radiation field for $10\,\mathrm{Myr}$, shown as the gray region. 
    The initially low accretion rate in the Pop III.1 scenario results from the hot, ionized gas produced by star. In this model, efficient BH accretion is already recovered $\sim 20\,\mathrm{Myr}$, and all memory of the seeding scenario is lost after $\sim 70\,\mathrm{Myr}$. In the presence of AGN feedback, we expect this period to be shorter and comparable to the onset of super Eddington accretion.}
    \label{fig:BH_acc}
\end{figure}

Figure~\ref{fig:BH_acc} shows the evolution of the black hole accretion rate following the formation of a SMBH from a Pop III.1 progenitor. The solid and dashed lines represent the Bondi-Hoyle and Eddington accretion rates, respectively. 
The accretion rate following BH formation remains sub-Eddington for the first few $\mathrm{Myr}$. 
This period of low accretion rate is due to the specific seed formation scenario (Sanati et. al. in prep.). The formation of a Pop III.1 star depletes a substantial fraction of the galaxy gas mass, and its subsequent radiative feedback heats and ionizes its surrounding gas, temporarily reducing the available cold gas supply for accretion. 
%
As a result, the SMBH seed forms in an initially hot and diffuse environment. This drives a short delay between BH formation and the onset of efficient accretion.
%
%
Despite this seeding scenario, we find accretion onto the SMBH to rapidly become efficient and match the Eddington limit, only after $\sim 20\,\mathrm{Myr}$. 
Even in the absence of AGN feedback, all memory of the seeding scenario is deleted from the growth rate in less than $\sim 100\,\mathrm{Myr}$.
In the presence of AGN feedback, we expect this memory to be erased more rapidly, by the time our SMBH reaches Eddington limit-like accretion rates. 

\bsp	
\label{lastpage}
\end{document}